\documentclass[
 reprint,
 amsmath,
 amssymb,
 aps,
]{revtex4-2}

\usepackage{graphicx}
\usepackage{dcolumn}
\usepackage{bm}
\usepackage{hyperref}
\usepackage{amsmath}
\usepackage{soul}
\usepackage{xcolor}

\renewcommand{\figurename}{Fig.}

\begin{document}

\title{Memory in neural activity: long-range order without criticality}

\author{Jay K.-C. Sun}
\thanks{These authors contributed equally.}
\author{Chesson Sipling}
\thanks{These authors contributed equally.}
\author{Yuan-Hang Zhang}
\thanks{These authors contributed equally.}

\author{Massimiliano Di Ventra}
\email{To whom correspondence should be addressed. Email: diventra@physics.ucsd.edu}

\affiliation{
 Department of Physics, University of California San Diego, La Jolla, CA, 92093-0319, USA
}




\maketitle

\section*{Abstract} 

The ``criticality hypothesis'', based on observed scale-free correlations in neural activity, posits that the brain operates at a critical point of transition between two phases. However, the validity of this hypothesis is still debated. Here, employing a commonly used model of cortical dynamics, we find that a {\it phase} of long-range order (LRO) in neural activity may be induced by memory (time non-locality) without invoking criticality. The cortical dynamics model contains fast and slow time scales that govern the neural and resource (memory) dynamics, respectively. When the resource dynamics are sufficiently slow, we observe an LRO phase, which manifests in avalanche size and duration probability distributions that are fit well by power laws. When the slow and fast time scales are no longer sufficiently distinct, LRO is destroyed. Since this LRO phase spans a wide range of parameters, it is robust against perturbations, unlike critical systems.





\section{Introduction}
\label{sec:intro}

Time non-locality, equivalent to {\it memory} in the context of this study, is a general property of physical systems, whether classical or quantum, wherein the system's response to a perturbation at a specific time is influenced not only by its current state but also by its history~\cite{Kubo_1957}. In other words, memory is a ubiquitous property of physical systems with important dynamical consequences. 

It is then natural to ask what the role of memory is in the neural activity of the biological brain. However, to the best of our knowledge, a systematic study of such a role has not been carried out yet. This is not a minor point since from seemingly unrelated research it was shown that memory can induce a long-range order phase in dynamical systems~\cite{sipling2025memoryinducedlongrangeorderdynamical}, a property exploited in a recent brain-inspired computing paradigm~\cite{memcomputing_book_di_ventra}. Similar conclusions have been reached in the study of networks of artificial spiking neurons currently explored for neuromorphic computing~\cite{zhang2024collectivedynamicslongrangeorder}.

Long-range order (LRO) means that the elementary units of a physical system correlate at long distances, much longer than the length of the inter-unit interaction. It is a typical property of critical points of continuous phase transitions, where scale-free distributions emerge~\cite{pathria2016statistical}. The connection to the dynamics of the biological brain is then very strong. This is because several experiments have demonstrated scale-free distributions in neural activities~\cite{Beggs_and_Plenz_avalanche, PhysRevLett.96.028107, brochini2016phase_support_criticality, scarpetta2018hysteresis_support_criticality}. These results have led to the suggestion, called the ``criticality hypothesis'', that the biological brain operates at the critical point of transition between two phases~\cite{Beggs_and_Plenz_avalanche, crit_hypothesis, chialvo2010emergent_support_criticality, o2022critical_support_criticality}. However, much debate has ensued as to the validity of this claim~\cite{beggs2012being_support_criticality, hesse2014self_support_criticality, wilting201925_perspective_article, PhysRevX.11.021059, beggs2022addressing}, with some work offering serious doubts~\cite{DestexheENEURO.0551-20.2021}, in part because LRO may emerge even in the absence of criticality~\cite{Chan2024.05.28.596196}. This result has also been confirmed by the studies we mentioned previously~\cite{sipling2025memoryinducedlongrangeorderdynamical, zhang2024collectivedynamicslongrangeorder}. In fact, if the criticality hypothesis were valid, the brain would be extremely sensitive to perturbations, as any critical point of a continuous phase transition is. On the other hand, if such LRO were a {\it phase} rather than a critical point, robustness against perturbations would be a natural byproduct.

In this work, we utilize a simple cortical dynamics model which is originally inspired by Wilson and Cowan~\cite{wilson_cowan} and further based upon the Landau-Ginzburg framework in di Santo {\it et al.}~\cite{munoz_paper}. We explicitly show, providing both analytical arguments and numerical experiments, that time non-locality (memory) induces a phase of LRO in neural activity, {\it not} a point of criticality. 

As we will introduce in more detail in Sec.~\ref{sec:model}, our mesoscopic cortical dynamics model contains two types of dynamical variables: neural activity and resource availability, defined pairwise at each point on a square lattice. The amount of available resources acts as the system's memory. Each type of variable can be characterized by a unique timescale. We find that when the memory degrees of freedom have a slower timescale than the neural dynamics, collective bursts or {\it avalanches} of neural activity are generated. The size and duration of these avalanches follow power-law distributions, meaning that arbitrarily large avalanches that couple much of the lattice are not so uncommon. This indicates that spatially distant regions of neural activity are strongly correlated; in other words, LRO exists. Importantly, this phase of LRO exists for a wide range of relative timescales between the neural and memory dynamics. Outside of this phase, we instead observe behavior consisting of ``down'' and ``up'' states. These correspond to regions of minimal and perpetual activity, respectively.

\begin{figure}[t]
    \centering
    \includegraphics[width=0.8\columnwidth]{figs/main_fig1_vertical_v4.pdf}
    \caption{An illustration of our phenomenological model. The bottom image depicts our 2D square lattice network, meant to emulate the cerebral cortex, where neural activities $\rho_{\vec{x}}$ are represented by blue cubes. Each cube has some amount of available resources $R_{\vec{x}}$, represented by the red filling and acting as memory degrees of freedom. The green lines represent diffusive couplings between any activity region and its four nearest-neighbors. The top image zooms in on one region; the orange arrows show the dynamical coupling between $\rho_{\vec{x}}$ and $R_{\vec{x}}$, and the flow fields $F_{\rho}$ and $F_{R}$ can be read from the right hand side of Eqs.~\ref{dynamics_eqns1} and~\ref{dynamics_eqns2}, respectively.}
    \label{fig:main_fig_1}
\end{figure}

\section{Model}
\label{sec:model}

Our mesoscopic cortical dynamics model is illustrated heuristically in Fig.~\ref{fig:main_fig_1}. At each site in a two-dimensional (2D) square lattice network of size $L^2$ exists one region of neural activity, the smallest functional unit in our model. These regions are constructed by applying a Landau-Ginzburg approach~\cite{stanley1971phase, binney1992theory} to the widely-accepted Wilson-Cowan model for neural activity~\cite{wilson_cowan}. Individual regions of activity $\rho_{\vec{x}}$, each consisting of thousands of individual neurons, are taken to follow a Wilson-Cowan equation. We take the region-dependent resource availability $R_{\vec{x}}$ dynamics to be similar to those of the continuous time Tsodyks-Markram model for synaptic plasticity~\cite{markram_resource_paper}. The activity regions are coupled diffusively, and stochasticity is introduced in the form of additive noise, both in line with the Landau-Ginzburg approach:

\begin{eqnarray}
&\dot{\rho}_{\vec{x}}(t) &= (-a + R_{\vec{x}})\rho_{\vec{x}} + b \rho_{\vec{x}}^2 - c \rho_{\vec{x}}^3 + h \nonumber \\
& &\hspace{3.5cm}+ D \nabla^2\rho_{\vec{x}} + \sigma \eta_{\vec{x}}\,,\quad\;\;\label{dynamics_eqns1} \\
&\dot{R}_{\vec{x}}(t) &=  \delta - \frac{1}{\tau_{D}} ( R_{\vec{x}}\rho_{\vec{x}} + \sigma\zeta_{\vec{x}} )\,,\quad\;\;\label{dynamics_eqns2} 
\end{eqnarray}


\noindent where ${\vec{x}} = (x_1, x_2)$ and $x_i = 1, 2, \dots, L$. In Eqs.~\ref{dynamics_eqns1} and~\ref{dynamics_eqns2}, $\eta_{\vec{x}}$ and $\zeta_{\vec{x}}$ both correspond to zero-mean, unit-variance Gaussian white noise.

The neural activity Eq.~\ref{dynamics_eqns1} aims to emulate the time evolution of the neural voltage. Diffusivity exists here in the form of $D \nabla^2\rho_{\vec{x}} = D \sum_{\vec{y} \in n.n.\vec{x}} (\rho_{\vec{y}} - \rho_{\vec{x}})$, where $n.n.\vec{x}$ represents the nearest-neighbors to a lattice position $\vec{x}$. Note that the lattice spacing has been normalized to $1$, so $D$ is a diffusive frequency. Following~\cite{munoz_paper}, we have chosen the following values for the parameters: $D=1$, $a=1$, $b=1.5$, $c=1$, $h=10^{-7}$, and $\sigma=0.1$. A detailed explanation of the remaining terms (in particular, the effects of coefficients $a$, $b$, $c$, and $h$) can be found in~\cite{munoz_paper}. The resource availability Eq.~\ref{dynamics_eqns2} holistically represents the available resources (e.g., ions, neurotransmitters, ATP, etc.) that may be consumed by the activity region at site $\vec{x}$. For our study, we consider the time scale $\tau_{D}$ as a control parameter, as it governs the rate of resource decay with respect to the fixed diffusive timescale $\sim 1/D$.

We emphasize that Eqs.~\ref{dynamics_eqns1} and~\ref{dynamics_eqns2} are very similar to the ones employed by di Santo {\it et al.}~\cite{munoz_paper} in their study of the emergence of scale-free avalanches at the edge of synchronization.  However, their study does not discuss or identify the role of memory in the generation of an LRO phase. In addition, to more directly focus on the role memory plays in inducing LRO, our dynamics feature a few changes from the model in~\cite{munoz_paper}. First, we introduce an identical source of Gaussian white noise in the resource dynamics, to better reflect the stochasticity of real physical systems. Second, our noise terms $\sigma \eta_{\vec{x}}$ and $\sigma \zeta_{\vec{x}}$ are additive, rather than multiplicative. Third, we represent the replenishing of resources by a constant parameter $\delta= 0.004$, rather than a function of $R_{\vec{x}}$. And fourth, we have decided to scale the noise term in the resource dynamics by the value of $\tau_{D}$, which ensures the noise does not dominate the resource dynamics when the product $R_{\vec{x}}\rho_{\vec{x}}$ is small and thus that $\tau_D$ is representative of a resource decay timescale for arbitrary noise strength $\sigma$. Despite these changes, we obtain phase structures similar to those reported in~\cite{munoz_paper} when other parameters are kept at the same values, giving us confidence that these modifications do not significantly influence our conclusions. 
In fact, as we will show in Sec.~\ref{sec:analytics}, the nearest-neighbor diffusive couplings, $D \nabla^2\rho_{\vec{x}}$, and the activity-resource couplings, $R_{\vec{x}}\rho_{\vec{x}}$, are the dominant factors in inducing LRO. Our numerical results in~\ref{sec:numerics} also support this conclusion.

\section{Analytical Understanding}
\label{sec:analytics}

As we have anticipated, we now provide analytical arguments that the resource-activity couplings and local activity diffusion are the key factors in inducing LRO in this system. For clarity, we make the following simplifications to our dynamics for the duration of this section. Importantly, we are not choosing to further alter the model itself, but to simply omit terms in our calculations that do not affect our conclusions, simplifying the theoretical analysis. With this in mind, we temporarily make the following simplifications to our model:

\begin{equation}
\left\{
\begin{aligned}
    \dot{\rho}_{\vec{x}}(t) \rightarrow \dot{\rho}_{\vec{x}}(t) &= R_{\vec{x}} \rho_{\vec{x}} + D \nabla^2 \rho_{\vec{x}}\;, \quad\quad\quad\quad\\
    \dot{R}_{\vec{x}}(t) \rightarrow \dot{R}_{\vec{x}}(t) &= - \frac{1}{\tau_D} R_{\vec{x}} \rho_{\vec{x}}\;,\quad\quad\quad\quad
\end{aligned}\right.
\end{equation}

\noindent namely, we consider only the activity-resource and diffusive coupling terms.

Consider the case where $\tau_D \gtrsim 1/D$, i.e., the timescale of resource decay is longer than the timescale of activity diffusion (which is true for archetypal spiking neural dynamics~\cite{wilson_cowan}). In this case, because the resources decay slowly, they act as a memory degree of freedom by preserving information about the state of the system from far in the past. We would like to demonstrate that, after a sufficiently long time, the presence of these activity-resource couplings induces effective activity-activity couplings that extend beyond nearest-neighbors. This is indicative of LRO in the activities.

To this end, we will perform a time coarse-graining approach inspired by~\cite{sipling2025memoryinducedlongrangeorderdynamical} to evaluate $\rho_{\vec{x}}(t)$ and $R_{\vec{x}}(t)$. Time is doubly discretized: first, coarsely over the timescale of the activity (memory) dynamics, $\tau_D$, and then more finely over the neural activity timescale $1/D$. Fig.~\ref{fig:main_fig_2} clarifies the discretization procedure. First, we evolve $\rho_{\vec{x}}(t)$ over one resource timescale $\tau_D$, so $R_{\vec{x}}(t)$ can be treated as approximately constant:

\begin{equation}
\begin{aligned}
    \rho_{\vec{x}}(\tau_D) & = \rho_{\vec{x}}(0) + \int_0^{\tau_D} dt \dot{\rho}_{\vec{x}}(t) \\
    & \approx \rho_{\vec{x}}(0) + \tau_D \big(R_{\vec{x}}(0) \overline{\rho_{\vec{x}, 0}} + D \nabla^2 \overline{\rho_{\vec{x}, 0}} \big).
\end{aligned}
\end{equation}

Using similar notation as in~\cite{sipling2025memoryinducedlongrangeorderdynamical}, we define $\overline{\rho_{\vec{x}, k}}$ as

\begin{equation}
    \overline{\rho_{\vec{x}, k}} \equiv \frac{1}{n} \sum_{p=0}^{n-1} \rho_{\vec{x}}(k \tau_D + p/D),
\end{equation}

\noindent which is just the discretized average value of the activity at site $\vec{x}$ over the interval $[k\tau_D, (k+1)\tau_D)$. Here, $n \equiv \lfloor \tau_D D \rfloor $ is an approximate ratio between characteristic timescales (or, equivalently, the number of activity timesteps in a single resource timestep, with $\lfloor \cdot \rfloor$ the floor function).
This averaging is justified in a similar manner to the $R_{\vec{x}}$ averaging: since the activity variables do not change much over timescales less than $1/D$, we can treat them as approximately constant during intervals of length $1/D$.

If we evaluate $\rho_{\vec{x}}(t)$ at some time $t > \tau_D$, the resources have had time to update, so their dynamics must now be considered. For example, when $t = 2 \tau_D$, we will first need to find $R_{\vec{x}}(\tau_D)$ using a similar method:

\begin{figure}[t]
    \centering
    \includegraphics[width=0.8\columnwidth]{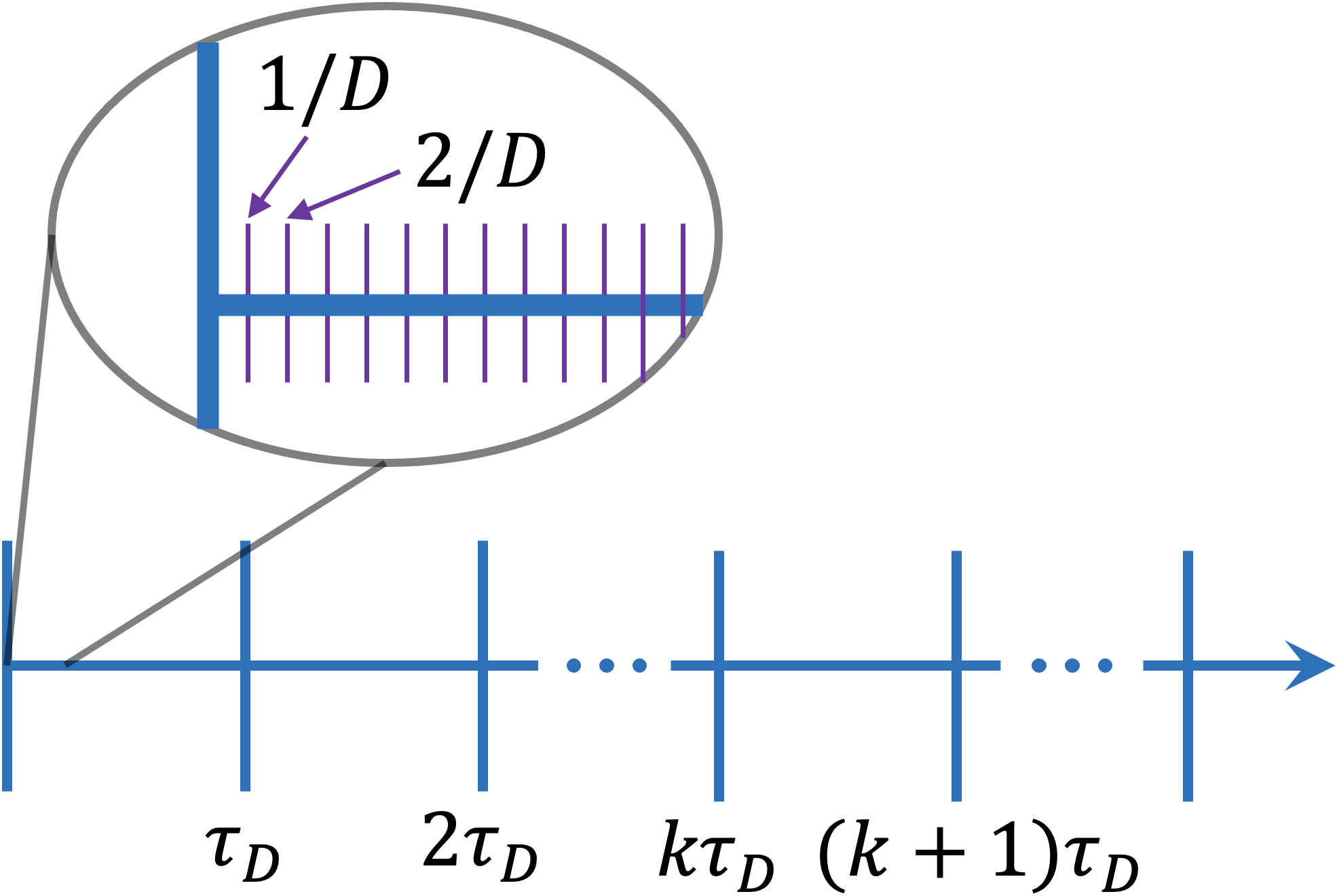}
    \caption{A visual depiction of the time interval over which our system evolves. We choose to discretize time into intervals of length $\tau_D$, the memory timescale, each of which is further discretized into intervals of length $1/D$, the neural activity timescale.}
    \label{fig:main_fig_2}
\end{figure}

\begin{equation}
\begin{aligned}
    R_{\vec{x}}(\tau_D) & = R_{\vec{x}}(0) + \int_0^{\tau_D} dt \dot{R}_{\vec{x}}(t) \\
    & \approx R_{\vec{x}}(0) (1 - \overline{\rho_{\vec{x}, 0}} ),
\end{aligned} 
\end{equation}

\noindent so that

\begin{equation}
\begin{aligned}
    \rho_{\vec{x}}(2 \tau_D) & \approx \rho_{\vec{x}}(\tau_D) + \tau_D \big( R_{\vec{x}}(\tau_D) \overline{\rho_{\vec{x}, 1}} + D \nabla^2 \overline{\rho_{\vec{x}, 1}} \big) \\
    & \approx \rho_{\vec{x}}(\tau_D) + \tau_D \big( R_{\vec{x}}(0) (1 - \overline{\rho_{\vec{x}, 0}}) \overline{\rho_{\vec{x}, 1}} + D \nabla^2 \overline{\rho_{\vec{x}, 1}} \big).
\end{aligned}
\end{equation}

Explicitly, we see that activity-activity couplings have been induced by the presence of memory in the term $- \tau_D R_{\vec{x}}(0) \overline{\rho_{\vec{x}, 0}} \, \overline{\rho_{\vec{x}, 1}}$. While this correlation appears to be only temporal, $\overline{\rho_{\vec{x}, 0}}$ and $\overline{\rho_{\vec{x}, 1}}$ implicitly depend on many of the sites around the lattice position $\vec{x}$ via diffusion. Since $\tau_D \gtrsim 1/D$, the diffusion will cause any $\rho_{\vec{x}}(\tau_D)$ to depend on initial activity values up to $n = \lfloor \tau_D D \rfloor$ lattice sites away.

Typically, this diffusive effect alone would decay exponentially and would not suggest LRO. However, because $\overline{\rho_{\vec{x}, 0}}$ feeds back into the dynamics for $\rho_{\vec{x}}(2 \tau_D)$ explicitly, beyond-nearest-neighbor couplings between activities in the lattice are created. These couplings are second order, but it can be shown that $N^{\text{th}}$ order couplings will be induced after time $N \tau_D$ has passed. We provide these generalizations along with more detailed justifications in Sec.~\ref{sec:SI_extended_analysis} of the Supplementary Information (SI). Therefore, after a sufficiently long time, we anticipate that regions of activity any arbitrary distance from one another in the lattice will become coupled.

\section{Numerical Results}
\label{sec:numerics}

\begin{figure*}[!htb]
    \centering
    \includegraphics[width=\linewidth]{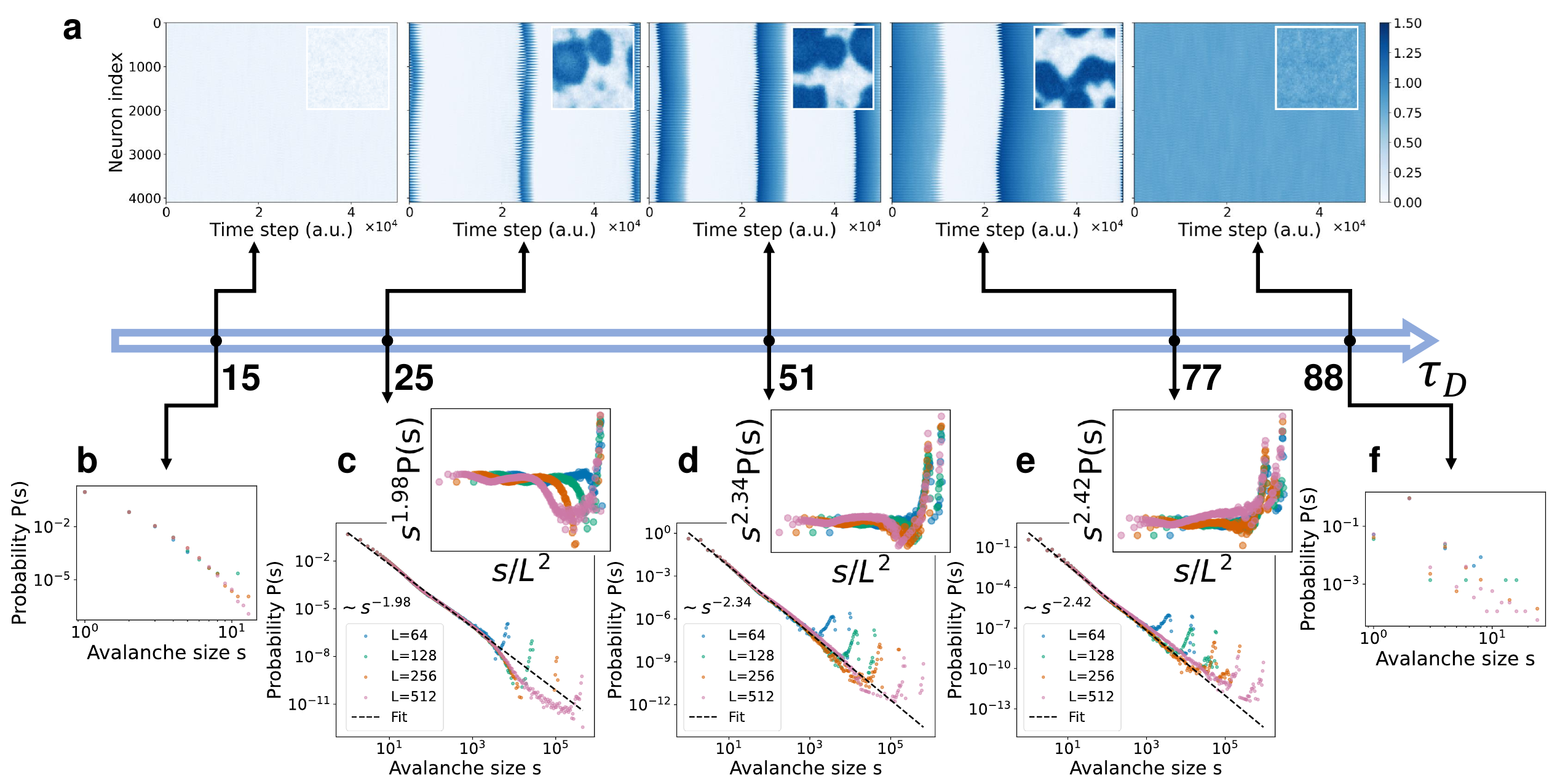}
    \caption{Illustration of the dynamics of neural activity $\rho$ and avalanche size distributions for different values of memory time scale $\tau_D$. (a) Raster plots of $\rho_{\vec{x}}(t)$ in a $N = 64^2$ lattice of activities. Each horizontal line represents the temporal evolution of one site, while the insets show corresponding 2D snapshots of spatial activity patterns at representative times. Videos of the temporal dynamics of this network for various $\tau_D$ can be found in the SI. As the value of $\tau_{D}$ increases, the system undergoes phase changes from the down phase (leftmost panel), to the LRO phase (middle 3 panels), to the rigid phase (rightmost panel). (b) Avalanche size distribution in the down phase ($\tau_D=15$). (c-e) Avalanche size distributions in the LRO phase ($\tau_D=25, 51, 77$). The insets are finite-size scaling curves transformed according to Eq.~\ref{decaying_power_law_distribution}, showing that the biases from ideal power-law distributions stem from finite-size effects. The additional peaks in the distributions in (d) and (e) correspond to avalanches that propagate back through the lattice more than once. (f) Avalanche size distribution in the rigid (up) phase ($\tau_D=88$). For $L=\{64, 128, 256, 512\}$, the statistics from each ensemble are taken from $\{100, 400, 2000, 4800\}$ instances, with each instance simulated for $\{50, 20, 5, 1.25\} \times 10^{4}$ time steps, respectively. Parameter values: $a = 1$, $b = 1.5$, $c = 1$, $h = 10^{-7}$, $D = 1$, $\sigma = 0.1$, $\delta = 0.004$, $\epsilon = 0.5$, $\Delta t=0.01$, and $\Delta_{tw}= 0.3$.}
    \label{fig:FIG_3}
\end{figure*}

To show explicitly the anticipated emergence of LRO in our cortical dynamics model, we simulate Eqs.~\ref{dynamics_eqns1} and~\ref{dynamics_eqns2} while varying the resource decay timescale, $\tau_D$. Fig.~\ref{fig:FIG_3}(a) depicts raster plots that visualize the neural activity $\rho(t)$ across the whole lattice (for N = $64^{2}$) for 5 different values of $\tau_{D}$, with the insets featuring 2D snapshots of the dynamics. Higher activity levels are indicated by darker blues. Each horizontal slice in the raster plot represents the activity level of a single region $\rho_{\vec{x}}(t)$. By varying $\tau_D$, with all other parameters fixed, three distinct signal patterns of $\rho(t)$ can be observed: a ``down'' phase, an intermediate phase, and an ``up'' phase.

When the resource depletion timescale is sufficiently short ($\tau_D \lesssim 25$, for our chosen parameters), the resources decay too fast for long-range couplings to be induced. This phase is characterized by a lack of spiking events altogether (small flares of neural activity persist due to the presence of noise). 

A rigid phase is observed when $\tau_{D}$ is large ($\tau_{D} \gtrsim 82$), meaning the resource decay rate is slow, allowing all regions to perpetually remain active. This is shown in the rightmost panel of Fig.~\ref{fig:FIG_3}(a). An intermediate phase emerges for $25 \lesssim \tau_D \lesssim 82$, characterized by spiking activity waves propagating through the entire lattice. This is depicted in the middle three panels in Fig.~\ref{fig:FIG_3}, where increasing $\tau_D$ increases the duration of the spiking activities. In this phase, we can expect more nuanced firing dynamics. 

\subsection{Avalanche Distributions}
\label{sec:distribution}

To further characterize the neural activity dynamics in each phase, we extract avalanche distributions from the neural activity levels $\rho_{\vec{x}}(t)$. See Sec.~\ref{sec:avalanche_definition} for more details on how we define and calculate avalanche size/duration distributions. We can then study how the distributions of avalanche sizes/durations change as $\tau_D$ is changed to further characterize each phase as a function of the resource depletion timescale. Avalanche size distributions are plotted in Fig.~\ref{fig:FIG_3}(b-f). Since the avalanche duration distributions look heuristically very similar to the avalanche size distributions, we choose to include the latter in Fig.~\ref{fig:big_size_fig_SI} of Sec.~\ref{sec:SI_size_distribution} of the SI. All following discussions of avalanche size/duration distributions can be applied interchangeably. All distributions are represented as histograms whose bin width is determined using Scott's normal reference rule~\cite{scott1979optimal}.

In the down phase ($\tau_D \lesssim 25$) and up phase ($\tau_D \gtrsim 82$), the activity level within the system remains approximately constant, with fluctuations induced by the noise term. As a consequence, $P(s)$, the probability of an avalanche with size $s$ occurring, quickly decays as $s$ increases, and the system is dominated by small avalanches. When $25 \lesssim \tau_D \lesssim 82$, we observe a power-law distribution of the avalanche sizes, with the largest avalanches spanning the entire system. This indicates that this is a phase with LRO.

Furthermore, we can corroborate our numerically detected exponents with the following well-known result from scaling theory~\cite{sethna_crackling}:

\begin{equation}
    \gamma = \frac{\alpha_T - 1}{\alpha_s - 1}\,.
    \label{eq:scaling_eqn}
\end{equation}

Above, $\alpha_s$ and $\alpha_T$ are the critical exponents for the avalanche size and duration distributions, respectively. $\gamma$ is defined via $\langle s \rangle(T) = T^\gamma$, in which, notably, the mean size of avalanches as a function of avalanche duration $T$ also follows a power law.

See Sec.~\ref{sec:scaling_relations} in the SI for a brief analytical explanation of this relation and our numerical distributions for $\langle s \rangle (T)$ over various $\tau_D$. Our distributions follow an approximate power-law which aligns well with Eq.~\ref{eq:scaling_eqn} for all $\tau_D$ within the LRO phase.

\subsection{Finite-Size Scaling}
\label{sec:finite_size_scaling}

We have also performed finite-size scaling analyses to confirm that, in the LRO phase, each avalanche distribution does approach a power-law decay distribution in the thermodynamic limit. If our distributions do abide by the finite-size scaling ansatz ~\cite{finite_size_scaling_Fisher}, then the deviations we see from strict power-law decays can be reasonably attributed to finite-size effects. This finite-size scaling ansatz imposes the following form on our avalanche probability distributions:

\begin{equation}\label{decaying_power_law_distribution}
\begin{aligned}
    P(v) \sim v^{-\alpha}\exp(-v/N^{\beta})\,,
\end{aligned}
\end{equation}

\noindent where $v$ =  $s$ or $T$ depending on the type of distribution we consider (spatial or temporal, respectively), N = $L^{2}$ is the system size, and parameters $\alpha$ and $\beta$ are the critical and cutoff exponents, respectively. The critical exponent $\alpha$ is the effective power-law exponent the distribution follows, whereas the cutoff exponent scales the distribution according to finite-size effects.

The scaling curves in the insets of Figs.~\ref{fig:FIG_3}(c-e) are obtained via transforming the size distribution curves according to Eq.~\ref{decaying_power_law_distribution}. The transformed vertical and horizontal axes are in units $s^{\alpha_s}P(s)$ and $s/N^{\beta_s}$, respectively. By carefully adjusting the scaling parameters $\alpha_s$ and $\beta_s$, we can collapse the distribution curves for all four system sizes fairly well. The collapsed curves in the insets of Figs.~\ref{fig:FIG_3}(c-e) confirm that the distribution curves do abide by the finite-size scaling law. This is most evident at $\tau_D=51$, while the rescaled curves start to deviate from perfect overlap as we approach the boundaries of the LRO phase. This suggests that the avalanche size distributions' departure from a power-law is, in fact, a finite-size effect; we then expect the distributions to approach power-law decays in the thermodynamic limit.

Similar finite-size scaling analysis holds for avalanche duration distributions, as shown in Fig.~\ref{fig:big_size_fig_SI} of SI Appendix Sec.~\ref{sec:SI_size_distribution}.

\subsection{Width of the LRO Phase and Correlation Length}
\label{sec:width_of_LRO}

\begin{figure}
    \centering
    \includegraphics[width=0.95\columnwidth]{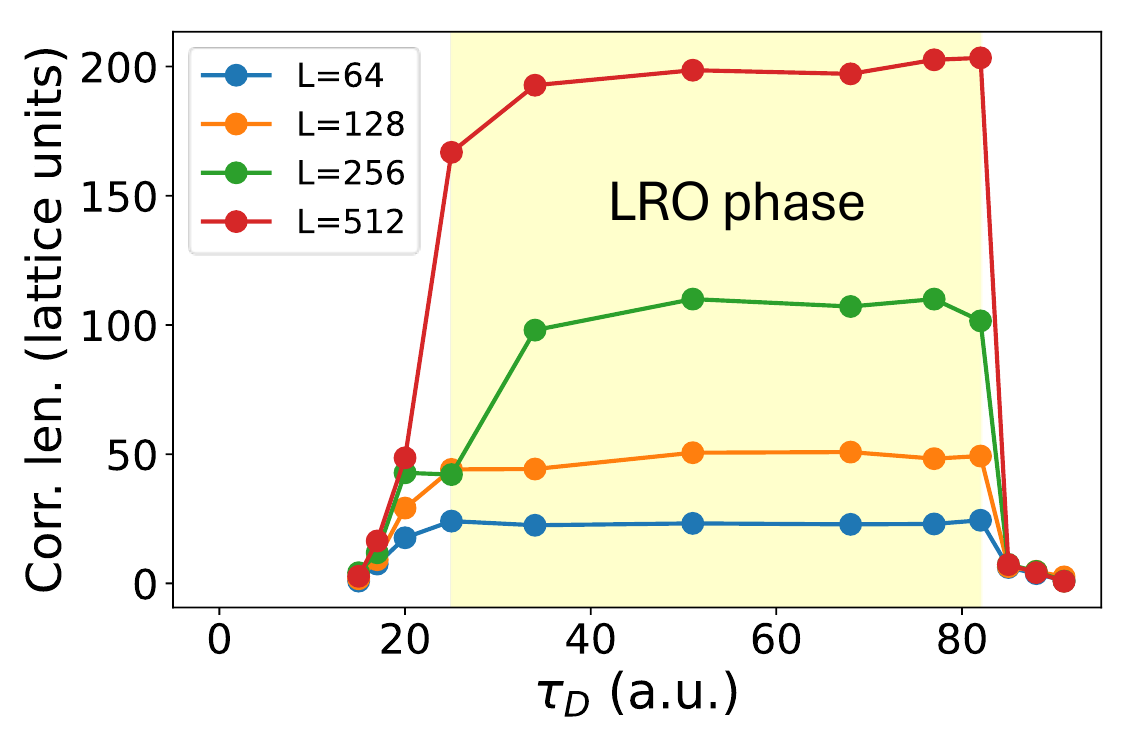}
    \caption{The correlation length, $\xi$, defined as the average distance between two units within the same avalanche (excluding system-wide avalanches), averaged over the entire distribution of avalanches \cite{christensen2002percolation}, as a function of $\tau_D$. For a large range of $\tau_D$ ($25 \lesssim \tau_D \lesssim 82$), the correlation length extends far beyond nearest-neighbors, indicating the existence of a robust phase of LRO.} 
    \label{fig:phase_diagram}
\end{figure}

Having computed the avalanche size distributions, we next define the correlation length, $\xi$, following the convention in percolation theory \cite{christensen2002percolation}. Specifically, we calculate the average distance between two units belonging to the same avalanche (excluding system-wide avalanches) and then average this quantity over the entire avalanche ensemble. The detailed formula for computing $\xi$ is provided in Sec.~\ref{sec:numerical_details}. This definition characterizes the typical spatial extent of an avalanche, which is inherently bounded by the system size $L$.

Fig.~\ref{fig:phase_diagram} shows $\xi$ as a function of $\tau_D$. Within the range $25 \leq \tau_D \leq 82$, $\xi(\tau_D) \gg 1$ (in units of lattice spacing), indicating correlations that extend well beyond nearest neighbors. In this regime, avalanches span a broad range of sizes, including events that approach the full lattice size, consistent with the emergence of long-range order (LRO). Moreover, $\xi$ increases systematically with $L$ in this phase, as larger lattices allow larger avalanches and therefore longer correlation lengths. This scaling behavior reflects the scale-free nature of the avalanche distributions: as $L$ increases, the upper cutoff of the power law extends, and in the thermodynamic limit ($L \to \infty$), $\xi$ is expected to diverge. This divergence signifies that the LRO phase is a genuine extended phase—robust across a wide parameter range—rather than a fine-tuned critical point.

\section{Discussion}
\label{sec:conclusion}

We have shown the key role memory (time non-locality) plays in influencing the collective dynamics of coupled neurons. More specifically, a separation between the characteristic neural activity (``spiking'') and resource decay timescales induces a  {\it phase} of long-range correlations in the neural activity that is not related to critical behavior. This phase is quite wide in terms of the memory decay timescale, suggesting robustness against perturbations.

A natural follow-up question is the following: what is the role that either the dimensionality or structure of the lattice plays in inducing LRO via memory in such a system?

We expect that the presence of such memory-induced LRO should generalize to higher dimensions as well as some alternative lattice structures, since our analytical understanding of this phenomenon does not explicitly depend on the lattice connectivity itself (cf. Sec.~\ref{sec:analytics}). On the contrary, this effect depends on (i) the existence of sufficiently strong memory (i.e., that the effective timescale of the memory is long) and (ii) that a sufficient number of pathways in the lattice exist whereby the memory effect can propagate through, correlating distant lattice sites. With {(ii) in mind, we anticipate that the {\it coordination number} (number of nearest-neighbors per site) is the relevant quantity which must be sufficiently large to give enough “pathways” for the memory effect to propagate. This, of course, depends on the lattice structure/dimension.

For example, since the coordination number is 6 for a 3D square lattice as opposed to 4 for the 2D case, we expect it to be easier for the memory-induced LRO state to form in a 3D lattice than a 2D one. On the other hand, LRO should not emerge in the 1D case. There, only a single path exists through which the memory effect can correlate any pair of sites. So, whenever the correlations become significant, {\it all} sites should become correlated, inducing a {\it rigid} state which is characterized by exclusively system-wide avalanches. Thus, there appears to be no way to obtain the power-law distribution of avalanches which characterizes LRO in a 1D lattice.

Additionally, we must emphasize that while similar phenomena are often attributed to self-organized criticality (SOC) \cite{bak1987self}, self-organized bistability (SOB) \cite{buendia2020self}, quasi-criticality \cite{bonachela2009self, buendia2020feedback, buendia2020self}, or the Griffiths phase \cite{moretti2013griffiths, vazquez2011temporal}, the observed memory-induced long-range order (LRO) in our system is distinct and cannot be fully captured within these existing frameworks.

Although criticality is strictly defined as a singular point, frameworks such as self-organized quasi-criticality (SOqC) \cite{bonachela2009self, buendia2020feedback}, self-organized quasi-bistability (SOqB) \cite{buendia2020feedback, buendia2020self, PhysRevResearch.3.023224}, and the Griffiths phase \cite{moretti2013griffiths, vazquez2011temporal} describe extended ranges of critical-like behavior. SOC requires the conservation of an effective ``energy'' (e.g., stress, resources), whereas SOqC and SOqB do not impose strict conservation laws \cite{bonachela2009self}. Instead, they rely on feedback mechanisms that drive the system toward a critical state, effectively ``smearing'' criticality and producing a range of critical-like behaviors \cite{buendia2020feedback}. However, our system does not rely on such mechanisms. 
Fig.~\ref{fig:phase_diagram} and Supplementary Figs.~\ref{fig:nullcline} and~\ref{fig:phase_diagram_noise} all show that the LRO regime occupies a broad region in parameter space, rather than lying solely near a region of phase transition. This suggests that it is not a mere extension of a critical point, but instead is a robust and expansive phase in its own right. While one could claim that is it merely the ``smearing'' of a critical point which gives the appearance of a phase, one would have to then prove the existence of such a critical point of a continuous/discontinuous phase transition (for SOC/SOB, respectively) in the first place, not a trivial task in itself. However, no such analysis regarding the parameter space of this cortical model is required if memory-induced LRO is the mechanism behind this behavior. Thus, we find this explanation more convincing, as it requires fewer assumptions than any type of quasi-criticality.

Analytically, as detailed in the SI Appendix Sec.~\ref{sec:SI_extended_analysis}, we show that this memory-induced LRO arises from long-range temporal interactions, which progressively and non-perturbatively correlate spatially distinct neurons. The requirement of non-perturbativeness enforces a separation of time scales between the primary and memory degrees of freedom. However, unlike in SOC and SOB, this separation does not need to diverge to infinity. A similar phenomenon of memory-induced LRO has been observed in neuromorphic systems \cite{zhang2024collectivedynamicslongrangeorder} and memory-coupled spin glasses \cite{sipling2025memoryinducedlongrangeorderdynamical}.

On the other hand, the Griffiths phase \cite{moretti2013griffiths, vazquez2011temporal} emerges from structural disorder, where noise induces exponentially rare large events that persist for exponentially long durations, leading to power-law distributions across a wide range of parameters. While noise is present in our model, its role is limited to breaking neural synchronization (see Supplementary Fig.~\ref{fig:phase_diagram_noise} and Supplementary Fig.~\ref{fig:kuramoto_index}), and large-scale events arise, at least in part, due to synchronization rather than noise-induced fluctuations. Furthermore, our relatively simple model lacks structural disorder, ruling out the possibility that our observed LRO phase corresponds to a Griffiths phase.


If this model were a faithful description of biological brain dynamics, these results would not support the criticality hypothesis. Irrespective, our work does reveal an alternative mechanism by which the brain could potentially exhibit long-range correlations between neurons without being in a critical state, namely through the action of {\it memory}. This is one of the fundamental properties of biological brains and its role in cortical dynamics should be explored further.

\section{Methods}
\label{sec:methods}

\subsection{Numerical Details}
\label{sec:numerical_details}

Eqs.~\ref{dynamics_eqns1} and~\ref{dynamics_eqns2} are numerically integrated using the Euler-Maruyama method with a fixed time step, $\Delta t=10^{-2}$. The Gaussian white noise is simulated as a Wiener process with zero mean and variance $\Delta t$ at each integration step. Initial conditions $\rho_{\vec{x}}(0)$ and $R_{\vec{x}}(0)$ were sampled from narrow Gaussian distributions centered at low values of $\rho_{\vec{x}}$ and $R_{\vec{x}}$ with variance 0.01. This choice minimizes the transient time while yielding the same characteristic (spiking) post-transient dynamics. Periodic boundary conditions are imposed at the edges of the lattice. Movies visualizing the dynamics for various values of $\tau_D$ are included in the SI. Lattice sizes $N \equiv L^2$ of $64^2$, $128^2$, $256^2$, and $512^2$ were simulated. We note that $\rho$ and $R$ are cut off at 0 from below, giving a positive bias to the otherwise Gaussian noise when $\rho$ and/or $R$ are small. For all simulations in the main text, we chose the parameters $a = 1$, $b = 1.5$, $c = 1$, $h = 10^{-7}$, $D = 1$, $\sigma = 0.1$, $\delta = 0.004$, $\epsilon = 0.5$, $\Delta t=0.01$, and $\Delta_{tw}= 0.3$ (generally in agreement with previous work~\cite{munoz_paper}) We study the effect of varying parameters $a$ and $\sigma$ in the SI.

The correlation length $\xi$, introduced in Fig.~\ref{fig:phase_diagram}, is defined as
\begin{equation}
    \xi^2=\frac{\sum_s 2R_s^2 s^2 n_s}{\sum_s s^2 n_s},
\end{equation}
where $s$ is the size of the avalanche, and $n_s$ is the number of avalanches of size $s$. Using the convention in percolation theory \cite{christensen2002percolation}, this summation excludes avalanches that span the entire system. $R_s$ is the radius of gyration of the avalanche, defined as the average distance to the center of mass \cite{christensen2002percolation}:
\begin{equation}
    R_s^2=\frac{1}{s}\sum_{i=1}^s |\mathbf{r}_i-\mathbf{r}_{cm}|^2.
\end{equation}
It can be shown that $R_s^2$ is half of the average square distance between two sites within the same avalanche \cite{christensen2002percolation}, $R_s^2=\frac{1}{2s^2}\sum_{ij}|\mathbf{r}_i-\mathbf{r_j}|^2$. Therefore, the correlation length $\xi$ represents the average distance of two sites belonging to the same avalanche.

\subsection{Neuronal Avalanches}
\label{sec:avalanche_definition}

In our simulated lattice of neural activity, we define avalanches as follows. When the activity at region $\vec{x}$, $\rho_{\vec{x}}(t)$, crosses a threshold $\epsilon$, we say an event has occurred at time $t$ and location $\vec{x}$. We then search for similar activity-crossing events at the four nearest-neighbors of $\vec{x}$ in the lattice during the following time $\Delta_{tw}$, called the ``time window''. Within the whole time window (from t to t + $\Delta_{tw}$), if an event is found at time $t'$, the size of the avalanche is increased by 1, and further events are searched for adjacent to the newest event between times $t'$ to $t' + \Delta_{tw}$. An avalanche stops when no events are found in any of the nearest-neighbors of recent (within a time $\Delta_{tw}$) events in that avalanche. As a result, a lone activity spike would characterize an avalanche of size 1. 

Note that our avalanche definition follows the approach in \cite{korchinski2021criticality}, which emphasizes spatial proximity and causality, rather than the experimentally inspired method that disregards spatial structure and relies solely on temporal binning \cite{Beggs_and_Plenz_avalanche, munoz_paper}. We argue that this physically motivated approach offers a more informative characterization of avalanche dynamics.

To yield sufficient avalanche statistics with limited computer memory, the dynamics are repeatedly simulated for a large number of times over an extended period for each lattice size $L^2$ and each $\tau_D$. To be specific, for $L=\{64, 128, 256, 512\}$, the number of repetitions are $\{100, 400, 2000, 4800\}$, and the duration of the simulations are $\{50, 20, 5, 1.25\}\times 10^4$ time steps, respectively, with each time step $\Delta t=0.01$.

\section*{Acknowledgments}
J.K.S. was supported by the George W. and Carol A. Lattimer Research Fellowship. Y.H.Z., C.S. and M.D. were supported by NSF grant No. ECCS-2229880.

\section*{Author Contributions}
M.D. suggested and supervised the work. J.K.S. and Y.H.Z. did the numerical simulations and data analysis. C.S. performed the theoretical analysis. All authors have read and contributed to the writing of the paper.

\section*{Competing Interests}
The authors declare no competing interests.

\section*{Data, Materials, and Software Availability}

The Python simulation code for data generation is available at~\cite{github_link}.

\bibliography{apssamp}

\begin{thebibliography}{39}%
\makeatletter
\providecommand \@ifxundefined [1]{%
 \@ifx{#1\undefined}
}%
\providecommand \@ifnum [1]{%
 \ifnum #1\expandafter \@firstoftwo
 \else \expandafter \@secondoftwo
 \fi
}%
\providecommand \@ifx [1]{%
 \ifx #1\expandafter \@firstoftwo
 \else \expandafter \@secondoftwo
 \fi
}%
\providecommand \natexlab [1]{#1}%
\providecommand \enquote  [1]{``#1''}%
\providecommand \bibnamefont  [1]{#1}%
\providecommand \bibfnamefont [1]{#1}%
\providecommand \citenamefont [1]{#1}%
\providecommand \href@noop [0]{\@secondoftwo}%
\providecommand \href [0]{\begingroup \@sanitize@url \@href}%
\providecommand \@href[1]{\@@startlink{#1}\@@href}%
\providecommand \@@href[1]{\endgroup#1\@@endlink}%
\providecommand \@sanitize@url [0]{\catcode `\\12\catcode `\$12\catcode `\&12\catcode `\#12\catcode `\^12\catcode `\_12\catcode `\%12\relax}%
\providecommand \@@startlink[1]{}%
\providecommand \@@endlink[0]{}%
\providecommand \url  [0]{\begingroup\@sanitize@url \@url }%
\providecommand \@url [1]{\endgroup\@href {#1}{\urlprefix }}%
\providecommand \urlprefix  [0]{URL }%
\providecommand \Eprint [0]{\href }%
\providecommand \doibase [0]{https://doi.org/}%
\providecommand \selectlanguage [0]{\@gobble}%
\providecommand \bibinfo  [0]{\@secondoftwo}%
\providecommand \bibfield  [0]{\@secondoftwo}%
\providecommand \translation [1]{[#1]}%
\providecommand \BibitemOpen [0]{}%
\providecommand \bibitemStop [0]{}%
\providecommand \bibitemNoStop [0]{.\EOS\space}%
\providecommand \EOS [0]{\spacefactor3000\relax}%
\providecommand \BibitemShut  [1]{\csname bibitem#1\endcsname}%
\let\auto@bib@innerbib\@empty
\bibitem [{\citenamefont {Kubo}(1957)}]{Kubo_1957}%
  \BibitemOpen
  \bibfield  {author} {\bibinfo {author} {\bibfnamefont {R.}~\bibnamefont {Kubo}},\ }\bibfield  {title} {\bibinfo {title} {{Statistical mechanical theory of irreversible processes. 1. General theory and simple applications in magnetic and conduction problems}},\ }\href {https://doi.org/10.1143/JPSJ.12.570} {\bibfield  {journal} {\bibinfo  {journal} {J. Phys. Soc. Jap.}\ }\textbf {\bibinfo {volume} {12}},\ \bibinfo {pages} {570} (\bibinfo {year} {1957})}\BibitemShut {NoStop}%
\bibitem [{\citenamefont {Sipling}\ \emph {et~al.}(2025)\citenamefont {Sipling}, \citenamefont {Zhang},\ and\ \citenamefont {Di~Ventra}}]{sipling2025memoryinducedlongrangeorderdynamical}%
  \BibitemOpen
  \bibfield  {author} {\bibinfo {author} {\bibfnamefont {C.}~\bibnamefont {Sipling}}, \bibinfo {author} {\bibfnamefont {Y.-H.}\ \bibnamefont {Zhang}},\ and\ \bibinfo {author} {\bibfnamefont {M.}~\bibnamefont {Di~Ventra}},\ }\bibfield  {title} {\bibinfo {title} {Memory-induced long-range order in dynamical systems},\ }\href {https://doi.org/10.1103/vwk9-79f7} {\bibfield  {journal} {\bibinfo  {journal} {Phys. Rev. E}\ }\textbf {\bibinfo {volume} {112}},\ \bibinfo {pages} {014124} (\bibinfo {year} {2025})}\BibitemShut {NoStop}%
\bibitem [{\citenamefont {Di~Ventra}(2022)}]{memcomputing_book_di_ventra}%
  \BibitemOpen
  \bibfield  {author} {\bibinfo {author} {\bibfnamefont {M.}~\bibnamefont {Di~Ventra}},\ }\href {https://doi.org/10.1093/oso/9780192845320.001.0001} {\emph {\bibinfo {title} {{MemComputing: Fundamentals and Applications}}}}\ (\bibinfo  {publisher} {Oxford University Press},\ \bibinfo {year} {2022})\BibitemShut {NoStop}%
\bibitem [{\citenamefont {Zhang}\ \emph {et~al.}(2024)\citenamefont {Zhang}, \citenamefont {Sipling}, \citenamefont {Qiu}, \citenamefont {Schuller},\ and\ \citenamefont {Di~Ventra}}]{zhang2024collectivedynamicslongrangeorder}%
  \BibitemOpen
  \bibfield  {author} {\bibinfo {author} {\bibfnamefont {Y.-H.}\ \bibnamefont {Zhang}}, \bibinfo {author} {\bibfnamefont {C.}~\bibnamefont {Sipling}}, \bibinfo {author} {\bibfnamefont {E.}~\bibnamefont {Qiu}}, \bibinfo {author} {\bibfnamefont {I.~K.}\ \bibnamefont {Schuller}},\ and\ \bibinfo {author} {\bibfnamefont {M.}~\bibnamefont {Di~Ventra}},\ }\bibfield  {title} {\bibinfo {title} {Collective dynamics and long-range order in thermal neuristor networks},\ }\href@noop {} {\bibfield  {journal} {\bibinfo  {journal} {Nature Communications}\ }\textbf {\bibinfo {volume} {15}},\ \bibinfo {pages} {6986} (\bibinfo {year} {2024})}\BibitemShut {NoStop}%
\bibitem [{\citenamefont {Pathria}(2016)}]{pathria2016statistical}%
  \BibitemOpen
  \bibfield  {author} {\bibinfo {author} {\bibfnamefont {R.~K.}\ \bibnamefont {Pathria}},\ }\href@noop {} {\emph {\bibinfo {title} {Statistical mechanics}}}\ (\bibinfo  {publisher} {Elsevier},\ \bibinfo {year} {2016})\BibitemShut {NoStop}%
\bibitem [{\citenamefont {Beggs}\ and\ \citenamefont {Plenz}(2003)}]{Beggs_and_Plenz_avalanche}%
  \BibitemOpen
  \bibfield  {author} {\bibinfo {author} {\bibfnamefont {J.~M.}\ \bibnamefont {Beggs}}\ and\ \bibinfo {author} {\bibfnamefont {D.}~\bibnamefont {Plenz}},\ }\bibfield  {title} {\bibinfo {title} {Neuronal avalanches in neocortical circuits},\ }\href {https://doi.org/10.1523/JNEUROSCI.23-35-11167.2003} {\bibfield  {journal} {\bibinfo  {journal} {Journal of Neuroscience}\ }\textbf {\bibinfo {volume} {23}},\ \bibinfo {pages} {11167} (\bibinfo {year} {2003})}\BibitemShut {NoStop}%
\bibitem [{\citenamefont {de~Arcangelis}\ \emph {et~al.}(2006)\citenamefont {de~Arcangelis}, \citenamefont {Perrone-Capano},\ and\ \citenamefont {Herrmann}}]{PhysRevLett.96.028107}%
  \BibitemOpen
  \bibfield  {author} {\bibinfo {author} {\bibfnamefont {L.}~\bibnamefont {de~Arcangelis}}, \bibinfo {author} {\bibfnamefont {C.}~\bibnamefont {Perrone-Capano}},\ and\ \bibinfo {author} {\bibfnamefont {H.~J.}\ \bibnamefont {Herrmann}},\ }\bibfield  {title} {\bibinfo {title} {Self-organized criticality model for brain plasticity},\ }\href {https://doi.org/10.1103/PhysRevLett.96.028107} {\bibfield  {journal} {\bibinfo  {journal} {Phys. Rev. Lett.}\ }\textbf {\bibinfo {volume} {96}},\ \bibinfo {pages} {028107} (\bibinfo {year} {2006})}\BibitemShut {NoStop}%
\bibitem [{\citenamefont {Brochini}\ \emph {et~al.}(2016)\citenamefont {Brochini}, \citenamefont {de~Andrade~Costa}, \citenamefont {Abadi}, \citenamefont {Roque}, \citenamefont {Stolfi},\ and\ \citenamefont {Kinouchi}}]{brochini2016phase_support_criticality}%
  \BibitemOpen
  \bibfield  {author} {\bibinfo {author} {\bibfnamefont {L.}~\bibnamefont {Brochini}}, \bibinfo {author} {\bibfnamefont {A.}~\bibnamefont {de~Andrade~Costa}}, \bibinfo {author} {\bibfnamefont {M.}~\bibnamefont {Abadi}}, \bibinfo {author} {\bibfnamefont {A.~C.}\ \bibnamefont {Roque}}, \bibinfo {author} {\bibfnamefont {J.}~\bibnamefont {Stolfi}},\ and\ \bibinfo {author} {\bibfnamefont {O.}~\bibnamefont {Kinouchi}},\ }\bibfield  {title} {\bibinfo {title} {Phase transitions and self-organized criticality in networks of stochastic spiking neurons},\ }\href@noop {} {\bibfield  {journal} {\bibinfo  {journal} {Scientific reports}\ }\textbf {\bibinfo {volume} {6}},\ \bibinfo {pages} {35831} (\bibinfo {year} {2016})}\BibitemShut {NoStop}%
\bibitem [{\citenamefont {Scarpetta}\ \emph {et~al.}(2018)\citenamefont {Scarpetta}, \citenamefont {Apicella}, \citenamefont {Minati},\ and\ \citenamefont {De~Candia}}]{scarpetta2018hysteresis_support_criticality}%
  \BibitemOpen
  \bibfield  {author} {\bibinfo {author} {\bibfnamefont {S.}~\bibnamefont {Scarpetta}}, \bibinfo {author} {\bibfnamefont {I.}~\bibnamefont {Apicella}}, \bibinfo {author} {\bibfnamefont {L.}~\bibnamefont {Minati}},\ and\ \bibinfo {author} {\bibfnamefont {A.}~\bibnamefont {De~Candia}},\ }\bibfield  {title} {\bibinfo {title} {Hysteresis, neural avalanches, and critical behavior near a first-order transition of a spiking neural network},\ }\href@noop {} {\bibfield  {journal} {\bibinfo  {journal} {Physical Review E}\ }\textbf {\bibinfo {volume} {97}},\ \bibinfo {pages} {062305} (\bibinfo {year} {2018})}\BibitemShut {NoStop}%
\bibitem [{\citenamefont {Beggs}(2008)}]{crit_hypothesis}%
  \BibitemOpen
  \bibfield  {author} {\bibinfo {author} {\bibfnamefont {J.~M.}\ \bibnamefont {Beggs}},\ }\bibfield  {title} {\bibinfo {title} {The criticality hypothesis: how local cortical networks might optimize information processing},\ }\href@noop {} {\bibfield  {journal} {\bibinfo  {journal} {Phil. Trans. R. Soc. A.}\ } (\bibinfo {year} {2008})}\BibitemShut {NoStop}%
\bibitem [{\citenamefont {Chialvo}(2010)}]{chialvo2010emergent_support_criticality}%
  \BibitemOpen
  \bibfield  {author} {\bibinfo {author} {\bibfnamefont {D.~R.}\ \bibnamefont {Chialvo}},\ }\bibfield  {title} {\bibinfo {title} {Emergent complex neural dynamics},\ }\href@noop {} {\bibfield  {journal} {\bibinfo  {journal} {Nature physics}\ }\textbf {\bibinfo {volume} {6}},\ \bibinfo {pages} {744} (\bibinfo {year} {2010})}\BibitemShut {NoStop}%
\bibitem [{\citenamefont {O’Byrne}\ and\ \citenamefont {Jerbi}(2022)}]{o2022critical_support_criticality}%
  \BibitemOpen
  \bibfield  {author} {\bibinfo {author} {\bibfnamefont {J.}~\bibnamefont {O’Byrne}}\ and\ \bibinfo {author} {\bibfnamefont {K.}~\bibnamefont {Jerbi}},\ }\bibfield  {title} {\bibinfo {title} {How critical is brain criticality?},\ }\href@noop {} {\bibfield  {journal} {\bibinfo  {journal} {Trends in Neurosciences}\ }\textbf {\bibinfo {volume} {45}},\ \bibinfo {pages} {820} (\bibinfo {year} {2022})}\BibitemShut {NoStop}%
\bibitem [{\citenamefont {Beggs}\ and\ \citenamefont {Timme}(2012)}]{beggs2012being_support_criticality}%
  \BibitemOpen
  \bibfield  {author} {\bibinfo {author} {\bibfnamefont {J.~M.}\ \bibnamefont {Beggs}}\ and\ \bibinfo {author} {\bibfnamefont {N.}~\bibnamefont {Timme}},\ }\bibfield  {title} {\bibinfo {title} {Being critical of criticality in the brain},\ }\href@noop {} {\bibfield  {journal} {\bibinfo  {journal} {Frontiers in physiology}\ }\textbf {\bibinfo {volume} {3}},\ \bibinfo {pages} {163} (\bibinfo {year} {2012})}\BibitemShut {NoStop}%
\bibitem [{\citenamefont {Hesse}\ and\ \citenamefont {Gross}(2014)}]{hesse2014self_support_criticality}%
  \BibitemOpen
  \bibfield  {author} {\bibinfo {author} {\bibfnamefont {J.}~\bibnamefont {Hesse}}\ and\ \bibinfo {author} {\bibfnamefont {T.}~\bibnamefont {Gross}},\ }\bibfield  {title} {\bibinfo {title} {Self-organized criticality as a fundamental property of neural systems},\ }\href@noop {} {\bibfield  {journal} {\bibinfo  {journal} {Frontiers in systems neuroscience}\ }\textbf {\bibinfo {volume} {8}},\ \bibinfo {pages} {166} (\bibinfo {year} {2014})}\BibitemShut {NoStop}%
\bibitem [{\citenamefont {Wilting}\ and\ \citenamefont {Priesemann}(2019)}]{wilting201925_perspective_article}%
  \BibitemOpen
  \bibfield  {author} {\bibinfo {author} {\bibfnamefont {J.}~\bibnamefont {Wilting}}\ and\ \bibinfo {author} {\bibfnamefont {V.}~\bibnamefont {Priesemann}},\ }\bibfield  {title} {\bibinfo {title} {25 years of criticality in neuroscience—established results, open controversies, novel concepts},\ }\href@noop {} {\bibfield  {journal} {\bibinfo  {journal} {Current opinion in neurobiology}\ }\textbf {\bibinfo {volume} {58}},\ \bibinfo {pages} {105} (\bibinfo {year} {2019})}\BibitemShut {NoStop}%
\bibitem [{\citenamefont {Korchinski}\ \emph {et~al.}(2021{\natexlab{a}})\citenamefont {Korchinski}, \citenamefont {Orlandi}, \citenamefont {Son},\ and\ \citenamefont {Davidsen}}]{PhysRevX.11.021059}%
  \BibitemOpen
  \bibfield  {author} {\bibinfo {author} {\bibfnamefont {D.~J.}\ \bibnamefont {Korchinski}}, \bibinfo {author} {\bibfnamefont {J.~G.}\ \bibnamefont {Orlandi}}, \bibinfo {author} {\bibfnamefont {S.-W.}\ \bibnamefont {Son}},\ and\ \bibinfo {author} {\bibfnamefont {J.}~\bibnamefont {Davidsen}},\ }\bibfield  {title} {\bibinfo {title} {Criticality in spreading processes without timescale separation and the critical brain hypothesis},\ }\href {https://doi.org/10.1103/PhysRevX.11.021059} {\bibfield  {journal} {\bibinfo  {journal} {Phys. Rev. X}\ }\textbf {\bibinfo {volume} {11}},\ \bibinfo {pages} {021059} (\bibinfo {year} {2021}{\natexlab{a}})}\BibitemShut {NoStop}%
\bibitem [{\citenamefont {Beggs}(2022)}]{beggs2022addressing}%
  \BibitemOpen
  \bibfield  {author} {\bibinfo {author} {\bibfnamefont {J.~M.}\ \bibnamefont {Beggs}},\ }\bibfield  {title} {\bibinfo {title} {Addressing skepticism of the critical brain hypothesis},\ }\href@noop {} {\bibfield  {journal} {\bibinfo  {journal} {Frontiers in computational neuroscience}\ }\textbf {\bibinfo {volume} {16}},\ \bibinfo {pages} {703865} (\bibinfo {year} {2022})}\BibitemShut {NoStop}%
\bibitem [{\citenamefont {Destexhe}\ and\ \citenamefont {Touboul}(2021)}]{DestexheENEURO.0551-20.2021}%
  \BibitemOpen
  \bibfield  {author} {\bibinfo {author} {\bibfnamefont {A.}~\bibnamefont {Destexhe}}\ and\ \bibinfo {author} {\bibfnamefont {J.~D.}\ \bibnamefont {Touboul}},\ }\bibfield  {title} {\bibinfo {title} {Is there sufficient evidence for criticality in cortical systems?},\ }\bibfield  {journal} {\bibinfo  {journal} {eNeuro}\ }\textbf {\bibinfo {volume} {8}},\ \href {https://doi.org/10.1523/ENEURO.0551-20.2021} {10.1523/ENEURO.0551-20.2021} (\bibinfo {year} {2021}),\ \Eprint {https://arxiv.org/abs/https://www.eneuro.org/content/8/2/ENEURO.0551-20.2021.full.pdf} {https://www.eneuro.org/content/8/2/ENEURO.0551-20.2021.full.pdf} \BibitemShut {NoStop}%
\bibitem [{\citenamefont {Chan}\ \emph {et~al.}(2024)\citenamefont {Chan}, \citenamefont {Kok},\ and\ \citenamefont {Ching}}]{Chan2024.05.28.596196}%
  \BibitemOpen
  \bibfield  {author} {\bibinfo {author} {\bibfnamefont {L.-C.}\ \bibnamefont {Chan}}, \bibinfo {author} {\bibfnamefont {T.-F.}\ \bibnamefont {Kok}},\ and\ \bibinfo {author} {\bibfnamefont {E.~S.}\ \bibnamefont {Ching}},\ }\bibfield  {title} {\bibinfo {title} {Emergence of a dynamical state of coherent bursting with power-law distributed avalanches from collective stochastic dynamics of adaptive neurons},\ }\bibfield  {journal} {\bibinfo  {journal} {bioRxiv}\ }\href {https://doi.org/10.1101/2024.05.28.596196} {10.1101/2024.05.28.596196} (\bibinfo {year} {2024})\BibitemShut {NoStop}%
\bibitem [{\citenamefont {Wilson}\ and\ \citenamefont {Cowan}(1972)}]{wilson_cowan}%
  \BibitemOpen
  \bibfield  {author} {\bibinfo {author} {\bibfnamefont {H.~R.}\ \bibnamefont {Wilson}}\ and\ \bibinfo {author} {\bibfnamefont {J.~D.}\ \bibnamefont {Cowan}},\ }\bibfield  {title} {\bibinfo {title} {Excitatory and inhibitory interactions in localized populations of model neurons},\ }\href@noop {} {\bibfield  {journal} {\bibinfo  {journal} {Biophysics Journal}\ }\textbf {\bibinfo {volume} {12}},\ \bibinfo {pages} {1} (\bibinfo {year} {1972})}\BibitemShut {NoStop}%
\bibitem [{\citenamefont {di~Santo}\ \emph {et~al.}(2018)\citenamefont {di~Santo}, \citenamefont {Villegas}, \citenamefont {Burioni},\ and\ \citenamefont {Muñoz}}]{munoz_paper}%
  \BibitemOpen
  \bibfield  {author} {\bibinfo {author} {\bibfnamefont {S.}~\bibnamefont {di~Santo}}, \bibinfo {author} {\bibfnamefont {P.}~\bibnamefont {Villegas}}, \bibinfo {author} {\bibfnamefont {R.}~\bibnamefont {Burioni}},\ and\ \bibinfo {author} {\bibfnamefont {M.~A.}\ \bibnamefont {Muñoz}},\ }\bibfield  {title} {\bibinfo {title} {{Landau–Ginzburg} theory of cortex dynamics: Scale-free avalanches emerge at the edge of synchronization},\ }\href {https://doi.org/10.1073/pnas.1712989115} {\bibfield  {journal} {\bibinfo  {journal} {Proceedings of the National Academy of Sciences}\ }\textbf {\bibinfo {volume} {115}},\ \bibinfo {pages} {E1356} (\bibinfo {year} {2018})}\BibitemShut {NoStop}%
\bibitem [{\citenamefont {Stanley}(1971)}]{stanley1971phase}%
  \BibitemOpen
  \bibfield  {author} {\bibinfo {author} {\bibfnamefont {H.~E.}\ \bibnamefont {Stanley}},\ }\href@noop {} {\emph {\bibinfo {title} {Phase transitions and critical phenomena}}},\ Vol.~\bibinfo {volume} {7}\ (\bibinfo  {publisher} {Clarendon Press, Oxford},\ \bibinfo {year} {1971})\BibitemShut {NoStop}%
\bibitem [{\citenamefont {Binney}\ \emph {et~al.}(1992)\citenamefont {Binney}, \citenamefont {Dowrick}, \citenamefont {Fisher},\ and\ \citenamefont {Newman}}]{binney1992theory}%
  \BibitemOpen
  \bibfield  {author} {\bibinfo {author} {\bibfnamefont {J.~J.}\ \bibnamefont {Binney}}, \bibinfo {author} {\bibfnamefont {N.~J.}\ \bibnamefont {Dowrick}}, \bibinfo {author} {\bibfnamefont {A.~J.}\ \bibnamefont {Fisher}},\ and\ \bibinfo {author} {\bibfnamefont {M.~E.}\ \bibnamefont {Newman}},\ }\href@noop {} {\emph {\bibinfo {title} {The theory of critical phenomena: an introduction to the renormalization group}}}\ (\bibinfo  {publisher} {Oxford University Press},\ \bibinfo {year} {1992})\BibitemShut {NoStop}%
\bibitem [{\citenamefont {Tsodyks}\ and\ \citenamefont {Markram}(1997)}]{markram_resource_paper}%
  \BibitemOpen
  \bibfield  {author} {\bibinfo {author} {\bibfnamefont {M.~V.}\ \bibnamefont {Tsodyks}}\ and\ \bibinfo {author} {\bibfnamefont {H.}~\bibnamefont {Markram}},\ }\bibfield  {title} {\bibinfo {title} {The neural code between neocortical pyramidal neurons depends on neurotransmitter release probability},\ }\href {https://doi.org/10.1073/pnas.94.2.719} {\bibfield  {journal} {\bibinfo  {journal} {Proceedings of the National Academy of Sciences}\ }\textbf {\bibinfo {volume} {94}},\ \bibinfo {pages} {719} (\bibinfo {year} {1997})}\BibitemShut {NoStop}%
\bibitem [{\citenamefont {Scott}(1979)}]{scott1979optimal}%
  \BibitemOpen
  \bibfield  {author} {\bibinfo {author} {\bibfnamefont {D.~W.}\ \bibnamefont {Scott}},\ }\bibfield  {title} {\bibinfo {title} {On optimal and data-based histograms},\ }\href@noop {} {\bibfield  {journal} {\bibinfo  {journal} {Biometrika}\ }\textbf {\bibinfo {volume} {66}},\ \bibinfo {pages} {605} (\bibinfo {year} {1979})}\BibitemShut {NoStop}%
\bibitem [{\citenamefont {Sethna}\ \emph {et~al.}(2001)\citenamefont {Sethna}, \citenamefont {Dahmen},\ and\ \citenamefont {Myers}}]{sethna_crackling}%
  \BibitemOpen
  \bibfield  {author} {\bibinfo {author} {\bibfnamefont {J.~P.}\ \bibnamefont {Sethna}}, \bibinfo {author} {\bibfnamefont {K.~A.}\ \bibnamefont {Dahmen}},\ and\ \bibinfo {author} {\bibfnamefont {C.~R.}\ \bibnamefont {Myers}},\ }\bibfield  {title} {\bibinfo {title} {Crackling noise},\ }\bibfield  {journal} {\bibinfo  {journal} {Nature}\ }\textbf {\bibinfo {volume} {410}},\ \href {https://doi.org/10.1038/35065675} {10.1038/35065675} (\bibinfo {year} {2001})\BibitemShut {NoStop}%
\bibitem [{\citenamefont {Fisher}\ and\ \citenamefont {Barber}(1972)}]{finite_size_scaling_Fisher}%
  \BibitemOpen
  \bibfield  {author} {\bibinfo {author} {\bibfnamefont {M.~E.}\ \bibnamefont {Fisher}}\ and\ \bibinfo {author} {\bibfnamefont {M.~N.}\ \bibnamefont {Barber}},\ }\bibfield  {title} {\bibinfo {title} {Scaling theory for finite-size effects in the critical region},\ }\href {https://doi.org/10.1103/PhysRevLett.28.1516} {\bibfield  {journal} {\bibinfo  {journal} {Phys. Rev. Lett.}\ }\textbf {\bibinfo {volume} {28}},\ \bibinfo {pages} {1516} (\bibinfo {year} {1972})}\BibitemShut {NoStop}%
\bibitem [{\citenamefont {Christensen}(2002)}]{christensen2002percolation}%
  \BibitemOpen
  \bibfield  {author} {\bibinfo {author} {\bibfnamefont {K.}~\bibnamefont {Christensen}},\ }\bibfield  {title} {\bibinfo {title} {Percolation theory},\ }\href@noop {} {\bibfield  {journal} {\bibinfo  {journal} {Imperial College London}\ }\textbf {\bibinfo {volume} {1}},\ \bibinfo {pages} {87} (\bibinfo {year} {2002})}\BibitemShut {NoStop}%
\bibitem [{\citenamefont {Bak}\ \emph {et~al.}(1987)\citenamefont {Bak}, \citenamefont {Tang},\ and\ \citenamefont {Wiesenfeld}}]{bak1987self}%
  \BibitemOpen
  \bibfield  {author} {\bibinfo {author} {\bibfnamefont {P.}~\bibnamefont {Bak}}, \bibinfo {author} {\bibfnamefont {C.}~\bibnamefont {Tang}},\ and\ \bibinfo {author} {\bibfnamefont {K.}~\bibnamefont {Wiesenfeld}},\ }\bibfield  {title} {\bibinfo {title} {Self-organized criticality: An explanation of the 1/f noise},\ }\href@noop {} {\bibfield  {journal} {\bibinfo  {journal} {Physical review letters}\ }\textbf {\bibinfo {volume} {59}},\ \bibinfo {pages} {381} (\bibinfo {year} {1987})}\BibitemShut {NoStop}%
\bibitem [{\citenamefont {Buend{\'\i}a}\ \emph {et~al.}(2020{\natexlab{a}})\citenamefont {Buend{\'\i}a}, \citenamefont {Di~Santo}, \citenamefont {Villegas}, \citenamefont {Burioni},\ and\ \citenamefont {Mu{\~n}oz}}]{buendia2020self}%
  \BibitemOpen
  \bibfield  {author} {\bibinfo {author} {\bibfnamefont {V.}~\bibnamefont {Buend{\'\i}a}}, \bibinfo {author} {\bibfnamefont {S.}~\bibnamefont {Di~Santo}}, \bibinfo {author} {\bibfnamefont {P.}~\bibnamefont {Villegas}}, \bibinfo {author} {\bibfnamefont {R.}~\bibnamefont {Burioni}},\ and\ \bibinfo {author} {\bibfnamefont {M.~A.}\ \bibnamefont {Mu{\~n}oz}},\ }\bibfield  {title} {\bibinfo {title} {Self-organized bistability and its possible relevance for brain dynamics},\ }\href@noop {} {\bibfield  {journal} {\bibinfo  {journal} {Physical Review Research}\ }\textbf {\bibinfo {volume} {2}},\ \bibinfo {pages} {013318} (\bibinfo {year} {2020}{\natexlab{a}})}\BibitemShut {NoStop}%
\bibitem [{\citenamefont {Bonachela}\ and\ \citenamefont {Munoz}(2009)}]{bonachela2009self}%
  \BibitemOpen
  \bibfield  {author} {\bibinfo {author} {\bibfnamefont {J.~A.}\ \bibnamefont {Bonachela}}\ and\ \bibinfo {author} {\bibfnamefont {M.~A.}\ \bibnamefont {Munoz}},\ }\bibfield  {title} {\bibinfo {title} {Self-organization without conservation: true or just apparent scale-invariance?},\ }\href@noop {} {\bibfield  {journal} {\bibinfo  {journal} {Journal of Statistical Mechanics: Theory and Experiment}\ }\textbf {\bibinfo {volume} {2009}},\ \bibinfo {pages} {P09009} (\bibinfo {year} {2009})}\BibitemShut {NoStop}%
\bibitem [{\citenamefont {Buend{\'\i}a}\ \emph {et~al.}(2020{\natexlab{b}})\citenamefont {Buend{\'\i}a}, \citenamefont {Di~Santo}, \citenamefont {Bonachela},\ and\ \citenamefont {Mu{\~n}oz}}]{buendia2020feedback}%
  \BibitemOpen
  \bibfield  {author} {\bibinfo {author} {\bibfnamefont {V.}~\bibnamefont {Buend{\'\i}a}}, \bibinfo {author} {\bibfnamefont {S.}~\bibnamefont {Di~Santo}}, \bibinfo {author} {\bibfnamefont {J.~A.}\ \bibnamefont {Bonachela}},\ and\ \bibinfo {author} {\bibfnamefont {M.~A.}\ \bibnamefont {Mu{\~n}oz}},\ }\bibfield  {title} {\bibinfo {title} {Feedback mechanisms for self-organization to the edge of a phase transition},\ }\href@noop {} {\bibfield  {journal} {\bibinfo  {journal} {Frontiers in physics}\ }\textbf {\bibinfo {volume} {8}},\ \bibinfo {pages} {333} (\bibinfo {year} {2020}{\natexlab{b}})}\BibitemShut {NoStop}%
\bibitem [{\citenamefont {Moretti}\ and\ \citenamefont {Mu{\~n}oz}(2013)}]{moretti2013griffiths}%
  \BibitemOpen
  \bibfield  {author} {\bibinfo {author} {\bibfnamefont {P.}~\bibnamefont {Moretti}}\ and\ \bibinfo {author} {\bibfnamefont {M.~A.}\ \bibnamefont {Mu{\~n}oz}},\ }\bibfield  {title} {\bibinfo {title} {Griffiths phases and the stretching of criticality in brain networks},\ }\href@noop {} {\bibfield  {journal} {\bibinfo  {journal} {Nature communications}\ }\textbf {\bibinfo {volume} {4}},\ \bibinfo {pages} {2521} (\bibinfo {year} {2013})}\BibitemShut {NoStop}%
\bibitem [{\citenamefont {Vazquez}\ \emph {et~al.}(2011)\citenamefont {Vazquez}, \citenamefont {Bonachela}, \citenamefont {L{\'o}pez},\ and\ \citenamefont {Munoz}}]{vazquez2011temporal}%
  \BibitemOpen
  \bibfield  {author} {\bibinfo {author} {\bibfnamefont {F.}~\bibnamefont {Vazquez}}, \bibinfo {author} {\bibfnamefont {J.~A.}\ \bibnamefont {Bonachela}}, \bibinfo {author} {\bibfnamefont {C.}~\bibnamefont {L{\'o}pez}},\ and\ \bibinfo {author} {\bibfnamefont {M.~A.}\ \bibnamefont {Munoz}},\ }\bibfield  {title} {\bibinfo {title} {Temporal griffiths phases},\ }\href@noop {} {\bibfield  {journal} {\bibinfo  {journal} {Physical review letters}\ }\textbf {\bibinfo {volume} {106}},\ \bibinfo {pages} {235702} (\bibinfo {year} {2011})}\BibitemShut {NoStop}%
\bibitem [{\citenamefont {Buend\'{\i}a}\ \emph {et~al.}(2021)\citenamefont {Buend\'{\i}a}, \citenamefont {Villegas}, \citenamefont {Burioni},\ and\ \citenamefont {Mu\~noz}}]{PhysRevResearch.3.023224}%
  \BibitemOpen
  \bibfield  {author} {\bibinfo {author} {\bibfnamefont {V.}~\bibnamefont {Buend\'{\i}a}}, \bibinfo {author} {\bibfnamefont {P.}~\bibnamefont {Villegas}}, \bibinfo {author} {\bibfnamefont {R.}~\bibnamefont {Burioni}},\ and\ \bibinfo {author} {\bibfnamefont {M.~A.}\ \bibnamefont {Mu\~noz}},\ }\bibfield  {title} {\bibinfo {title} {Hybrid-type synchronization transitions: Where incipient oscillations, scale-free avalanches, and bistability live together},\ }\href {https://doi.org/10.1103/PhysRevResearch.3.023224} {\bibfield  {journal} {\bibinfo  {journal} {Phys. Rev. Res.}\ }\textbf {\bibinfo {volume} {3}},\ \bibinfo {pages} {023224} (\bibinfo {year} {2021})}\BibitemShut {NoStop}%
\bibitem [{\citenamefont {Korchinski}\ \emph {et~al.}(2021{\natexlab{b}})\citenamefont {Korchinski}, \citenamefont {Orlandi}, \citenamefont {Son},\ and\ \citenamefont {Davidsen}}]{korchinski2021criticality}%
  \BibitemOpen
  \bibfield  {author} {\bibinfo {author} {\bibfnamefont {D.~J.}\ \bibnamefont {Korchinski}}, \bibinfo {author} {\bibfnamefont {J.~G.}\ \bibnamefont {Orlandi}}, \bibinfo {author} {\bibfnamefont {S.-W.}\ \bibnamefont {Son}},\ and\ \bibinfo {author} {\bibfnamefont {J.}~\bibnamefont {Davidsen}},\ }\bibfield  {title} {\bibinfo {title} {Criticality in spreading processes without timescale separation and the critical brain hypothesis},\ }\href@noop {} {\bibfield  {journal} {\bibinfo  {journal} {Physical Review X}\ }\textbf {\bibinfo {volume} {11}},\ \bibinfo {pages} {021059} (\bibinfo {year} {2021}{\natexlab{b}})}\BibitemShut {NoStop}%
\bibitem [{git()}]{github_link}%
  \BibitemOpen
  \href@noop {} {}\bibinfo {note} {\url{https://github.com/JaySun1207/memory_induced_LRO}}\BibitemShut {NoStop}%
\bibitem [{\citenamefont {Sethna}\ \emph {et~al.}(2005)\citenamefont {Sethna}, \citenamefont {Dahmen},\ and\ \citenamefont {Perkovic}}]{sethna2005randomfieldisingmodelshysteresis}%
  \BibitemOpen
  \bibfield  {author} {\bibinfo {author} {\bibfnamefont {J.~P.}\ \bibnamefont {Sethna}}, \bibinfo {author} {\bibfnamefont {K.~A.}\ \bibnamefont {Dahmen}},\ and\ \bibinfo {author} {\bibfnamefont {O.}~\bibnamefont {Perkovic}},\ }\href {https://arxiv.org/abs/cond-mat/0406320} {\bibinfo {title} {Random-field ising models of hysteresis}} (\bibinfo {year} {2005}),\ \Eprint {https://arxiv.org/abs/cond-mat/0406320} {arXiv:cond-mat/0406320 [cond-mat.mtrl-sci]} \BibitemShut {NoStop}%
\bibitem [{\citenamefont {Strogatz}(2003)}]{strogatz2003synchronization}%
  \BibitemOpen
  \bibfield  {author} {\bibinfo {author} {\bibfnamefont {S.}~\bibnamefont {Strogatz}},\ }\bibfield  {title} {\bibinfo {title} {Synchronization: a universal concept in nonlinear sciences},\ }\href@noop {} {\bibfield  {journal} {\bibinfo  {journal} {Physics Today}\ }\textbf {\bibinfo {volume} {56}},\ \bibinfo {pages} {47} (\bibinfo {year} {2003})}\BibitemShut {NoStop}%
\end{thebibliography}%
\clearpage
\newpage
\appendix

\begin{center}
{\large \bf Supplementary Information: Memory in neural activity: long-range order without criticality}
\end{center} 

\renewcommand{\figurename}{Supplementary Fig.}
\setcounter{figure}{0}

\setcounter{page}{1}
\setcounter{equation}{0} 

\section{Generalized Coarse-Graining Approach}
\label{sec:SI_extended_analysis}

Here, we return to our simplified cortical dynamics model, continuing to emphasize the importance of resource-activity and diffusive couplings:

\begin{equation}
\left\{
\begin{aligned}
    \dot{\rho}_{\vec{x}}(t) &= R_{\vec{x}} \rho_{\vec{x}} + D \nabla^2 \rho_{\vec{x}}\;, \\
    \dot{R}_{\vec{x}}(t) &= - \frac{1}{\tau_D} R_{\vec{x}} \rho_{\vec{x}}\;.
\end{aligned}\right.
\end{equation}

Let us first generalize our results to the case where $t = N \tau_D$, for arbitrarily large $N$:

\begin{equation}
\begin{aligned}
    R_{\vec{x}}(N \tau_D) & = R_{\vec{x}} \big( (N-1) \tau_D \big) + \int_{(N-1) \tau_D}^{N \tau_D} dt \dot{R}_{\vec{x}}(t) \\
    & \approx R_{\vec{x}} \big( (N-1) \tau_D \big) (1 - \overline{\rho_{\vec{x}, N-1}} ) \\
    & \approx R_{\vec{x}} \big( (N-2) \tau_D \big) (1 - \overline{\rho_{\vec{x}, N-2}} ) (1 - \overline{\rho_{\vec{x}, N-1}} ) \\
    & \approx R_{\vec{x}}(0) \prod_{l = 0}^{N-1} (1 - \overline{\rho_{\vec{x}, l}} )\;,
\end{aligned} 
\end{equation}

\begin{equation}
\begin{aligned}
    \rho_{\vec{x}}(N &\tau_D) \\
    &\approx \rho_{\vec{x}} \big( (N - 1) \tau_D \big) + \int_{(N-1) \tau_D}^{N \tau_D} dt \dot{\rho}_{\vec{x}}(t) \\
    &\approx \rho_{\vec{x}} \big( (N - 1) \tau_D \big) \\
    &\quad + \tau_D \bigg( R_{\vec{x}} \big( (N - 1) \tau_D \big) \overline{\rho_{\vec{x}, N-1}} + D \nabla^2 \overline{\rho_{\vec{x}, N-1}} \bigg) \\ 
    &\approx \rho_{\vec{x}} \big( (N - 1) \tau_D \big) \\
    &\quad + \tau_D \bigg( R_{\vec{x}}(0) \overline{\rho_{\vec{x}, N-1}} \prod_{l = 0}^{N - 2} (1 - \overline{\rho_{\vec{x}, l}}) + D \nabla^2 \overline{\rho_{\vec{x}, N-1}} \bigg) \\
    &\approx \rho_{\vec{x}}(0) \\ 
    &\quad + \tau_D \sum_{m = 0}^{N - 1} \bigg( R_{\vec{x}}(0) \overline{\rho_{\vec{x}, m}} \prod_{l = 0}^{m - 1} (1 - \overline{\rho_{\vec{x}, l}}) + D \nabla^2 \overline{\rho_{\vec{x}, m}} \bigg).
\end{aligned}
\end{equation}

For $m=0$, replace $\prod_{l=0}^{m-1} (1 - \overline{\rho_{\vec{x}, l}})$ with $1$. Above, we extrapolate the forms of $R_{\vec{x}}(N \tau_D)$ and $\rho_{\vec{x}}(N \tau_D)$ by determining their forms in terms of $R_{\vec{x}}((N - 1) \tau_D)$ and $\rho_{\vec{x}}((N - 1) \tau_D)$, and then iterating.

We see that the highest order activity-activity coupling terms are of the following form:

\begin{equation}
\label{coupling_term}
    \tau_D R_{\vec{x}}(0) \overline{\rho_{\vec{x}, m}} \prod_{l=0}^{m-1} (-\overline{\rho_{\vec{x}, l}})\,,
\end{equation}

\noindent where $m = 1, \dots, N - 1$ ($m = 0$ corresponds to no activity-activity coupling).

Let us now more explicitly investigate the discrete average $\overline{\rho_{\vec{x}, k}}$ that was introduced in the main text. First, we study $\rho_{\vec{x}}(k \tau_D + p/D)$, the constituent terms in its summation. For the case of $k = 0$,

\begin{equation}
\begin{aligned}
    \rho_{\vec{x}}(1/D) &= \rho_{\vec{x}}(0) + \int_{0}^{1/D} dt \dot{\rho}_{\vec{x}}(t) \\
    & \approx \rho_{\vec{x}}(0) + \frac{1}{D} \bigg( R_{\vec{x}}(0) \rho_{\vec{x}}(0) + D \nabla^2 \rho_{\vec{x}}(0) \bigg) \\
    & = \rho_{\vec{x}}(0) \big( 1 + R_{\vec{x}}(0)/D \big) + \sum_{\vec{y} \in n.n.\vec{x}} \bigg( \rho_{\vec{y}}(0) - \rho_{\vec{x}}(0) \bigg)\,, \\
    \rho_{\vec{x}}(2/D) & = \rho_{\vec{x}}(1/D) + \int_{1/D}^{2/D} dt \dot{\rho}_{\vec{x}}(t) \\
    & \approx \rho_{\vec{x}}(1/D) + \frac{1}{D} \bigg( R_{\vec{x}}(0) \rho_{\vec{x}}(1/D) + D \nabla^2 \rho_{\vec{x}}(1/D) \bigg) \\
    & = \cdots + \sum_{\vec{y} \in n.n.\vec{x}}\Bigg[ \, \sum_{\vec{z} \in n.n.\vec{y}} \bigg( \rho_{\vec{z}}(0) - \rho_{\vec{y}}(0) \bigg) \\
    &\quad\quad\quad\quad - \sum_{\vec{y} \in n.n.\vec{x}} \bigg( \rho_{\vec{y}}(0) - \rho_{\vec{x}}(0) \bigg) \Bigg]\,.
\end{aligned}
\end{equation}

Therefore, after 2 activity timesteps, because each neighbor at site $\vec{x}$ will have had time to
update due to local interactions with its neighbors, the activity at site $\vec{x}$ will depend on all the neighbors of the activities at $\vec{y}$ as well (denoted here with $\vec{z}$). This pattern will continue so that over a timescale $\tau_D$, each activity will implicitly depend on all its neighbors within $n$ lattice sites away (again, $n \equiv \lfloor \tau_D D \rfloor $ is the ratio of the resource to activity timescales). This occurs at every point in the lattice simultaneously.


This further generalizes for $k \neq 0$ (times longer than one resource timescale). In those instances, the form of $\rho_{\vec{x}}(k \tau_D + p/D)$ will be even more complicated, as the value of $R_{\vec{x}}(t)$ will vary during the dynamics. However, diffusive coupling will still cause individual activities to implicitly depend on the state of other activities farther and farther away. In this case, the radius over which implicit dependencies exist is $n N = N \lfloor \tau_D D \rfloor$, which is proportional to the total number of resource timesteps that have passed and the ratio between the activity and resource timescales.

To clarify, it is {\it not} this exponentially weakening radius of interactions due to the diffusive coupling that we interpret as LRO. Let us return to Eq.~\ref{coupling_term}. Now, it is clear that $\overline{\rho_{\vec{x}, k}}$ depends on the state of other activities within a wide radius of influence, albeit weakly. So, despite the contribution of individual activity regions being exponentially weak, this product series term suggests true, long-range activity-activity coupling up to order $N$, as each $\overline{\rho_{\vec{x}, k}}$ depends implicitly on activies far away. This makes sense, as it is the resource (memory) variables that induce this LRO, which ``update'' $N$ times over a duration of time $N \tau_D$.

Note that this argument relies solely on the coupling of neural activities to resources (and vice versa), {\it local} diffusive couplings in the activies, and that the ratio of the relevant timescales is sufficiently large.

Although this analysis does not imply {\it precisely} how large $\tau_D D$ must be to induce LRO, our numerical results show that the dynamics become rigid when $\tau_D \gtrsim 82$ (recall that we choose $D = 1$). Furthermore, the existence of a lower bound of the LRO phase ($\tau_D \lesssim 25$ corresponding to the down phase) is not evident from our previous considerations. As we explain in the next section, additional nuances in our cortical dynamics model prevent us from predicting the exact range of $\tau_D$ over which we expect LRO behavior from this analysis alone (previously ignored terms will play a role in determining the strength of emergent long-range couplings). However, the fundamental logic, that time non-locality can induce long-range coupling in locally coupled variables, is model-independent.

\begin{figure*}[htbp]
    \centering
    \includegraphics[width=\linewidth]{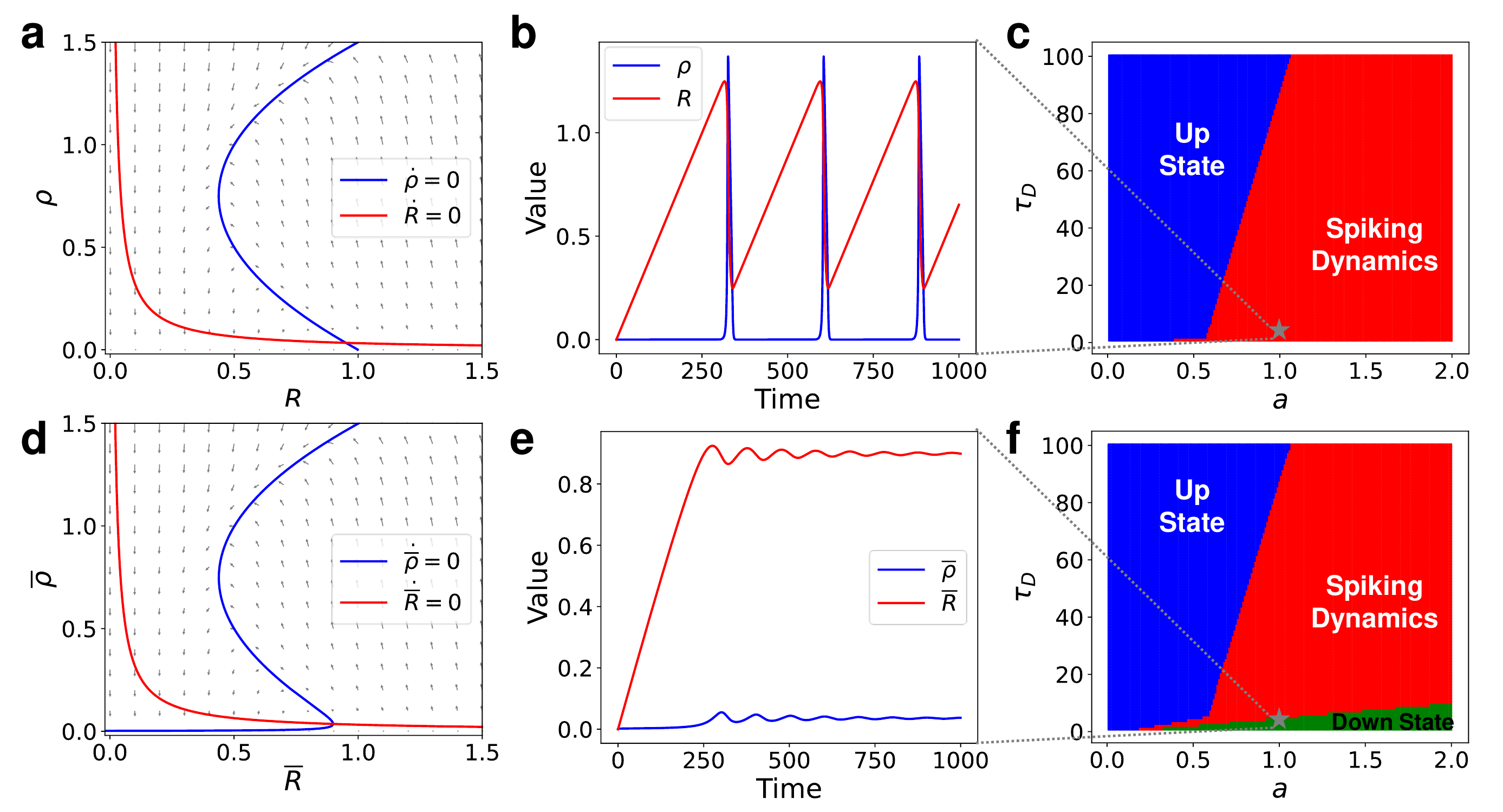}
    \caption{(a, d) The nullclines of the dynamics, for (a) a single neural activity region (from the deterministic, noiseless versions of Eqs.~\ref{dynamics_eqns1} and~\ref{dynamics_eqns2}), and (d) the mean field equations, Eq.~\ref{eq:mean}, both at $\tau_D=8$, $a=1$. The flow field is depicted with gray arrows. For better visualization, the length of each arrow is proportional to the logarithm of the flow field's magnitude. The intersection point is unstable in panel (a), but stable in panel (d), revealing that the existence of the down phase relies on the collective dynamics within the system. (b, e) The dynamics of (b) a single neural activity region and (e) the mean fields, both at $\tau_D=8$, $a=1$. We see a stable spiking oscillation in panel (b), while there is no oscillation in panel (e), corresponding to the down phase. (c)(f) Phase diagrams of (c) a single activity region, and (f) the mean field dynamics, with $\tau_D$ and $a$ as parameters. While no down state exists for an isolated activity region, a narrow stripe of the down phase emerges at low $\tau_D$ in the mean-field dynamics. Additionally, the parameter range of the LRO phase reported in the main text ($a = 1$, $25 \lesssim \tau_D \lesssim 82$) does not lie near any regions of phase transition, ruling out the possibility of criticality or quasi-criticality.
    }
    \label{fig:nullcline}
\end{figure*}

\section{Additional Numerical Results}

\subsection{Phase Structure of Cortical Dynamics}

We have previously identified three distinct phases in our system through numerical analysis: a down phase characterized by an absence of neural activities, an LRO phase marked by periodic spiking and extensive long-range correlations, and a rigid phase dominated by continuous neural activities. The previous section provided a framework for how the coupling of $\rho$ and $R$, which have different characteristic time scales, could produce LRO. In this section, we aim to semi-analytically determine the upper and lower bounds of this LRO phase. We find reasonably good agreement with numerical results.

Briefly, we study the phase structure of a single (deterministic) region of neural activity under these dynamics (i.e., with diffusive and noise terms removed from Eqs.~\ref{dynamics_eqns1} and~\ref{dynamics_eqns2}). As we will see, there is a simple interpretation for the transition between the LRO and up phases. 

First, we calculate the nullclines  $R_\rho(\rho)$ and $R_R(\rho)$ (curves with $\dot{\rho} = 0$ and $\dot{R} = 0$, respectively) of these dynamics to determine the nature of the fixed points $(R^*, \rho^*)$ for a given $\tau_D$. From solving deterministic, noiseless versions of Eqs.~\ref{dynamics_eqns1} and~\ref{dynamics_eqns2} for $R$, we find

\begin{equation}\label{nullclines}
\begin{aligned}
    R_\rho(\rho) &= c \rho^2 - b \rho + a - h/\rho\,,\\
    R_R(\rho) &= \delta \tau_D / \rho\,.
\end{aligned}
\end{equation}

We plot the flow field $(\dot{R}, \dot{\rho})$ in Fig.~\ref{fig:nullcline}(a), with nullclines $R_\rho(\rho)$ and $R_R(\rho)$ drawn in blue and red, respectively. 
Looking at the flow vector fields, note that $\tau_D$ influences the point at which the nullclines intersect, thereby determining the stability of the critical point $(R^*, \rho^*)$. This point transitions from a saddle point to a stable point when $\rho^*$ exceeds $\rho_{\text{min}}$, the minimum of $R_\rho$. This minimum occurs at $\rho_{\text{min}} \approx b/2c = 0.75$ (treating $R_\rho(\rho)$ as a quadratic function, since $h = 10^{-7} \approx 0$, and using our chosen parameters).

This splits the single-unit dynamics into two regimes based on the resource decay timescale $\tau_D$: one in which the dynamics quickly move towards a fixed point with constant $\rho$ and $R$, and one with a hysteretic loop corresponding to spiking dynamics. Thus, this analysis suggests that individual units cannot support spiking dynamics when $R_R(\rho_{\text{min}}) > R_\rho(\rho_{\text{min}})$. From this relationship and Eqs.~\ref{nullclines}, we obtain the following constraint on $\tau_D$ which permits spiking dynamics:

\begin{equation}\label{LROUpperBound}
    \tau_D < \frac{b}{2 c \delta} \bigg( \frac{b^2}{4 c} - \frac{b^2}{2 c} + a \bigg).
\end{equation}

For our chosen parameters, this gives an upper bound for the LRO phase of $\tau_D \approx 82.0$, which is in very good agreement with the numerical resuls presented in Sec.~\ref{sec:width_of_LRO}, see Fig.~\ref{fig:phase_diagram}.

However, this approach does not explain the existence of the down phase. Notably, the down phase does not occur if only one region of neural activity is considered. As illustrated in Fig.~\ref{fig:nullcline}(a), the intersection of the two nullclines remains unstable for all $\rho < b/2c=0.75$, which invariably results in periodic spiking dynamics, see Fig.~\ref{fig:nullcline}(b).

Instead, noise and diffusion are crucial factors in the down phase. Considering Eqs.~\ref{dynamics_eqns1} and~\ref{dynamics_eqns2} on an $L\times L$ lattice with periodic boundary condition (PBC), we can derive the dynamics of the mean fields, $\overline{\rho}=\sum_{\vec{x}}\rho_{\vec{x}}/L^2$ and $\overline{R}=\sum_{\vec{x}} R_{\vec{x}}/L^2$, by averaging over the entire lattice:

\begin{equation}
\begin{aligned}
    \dot{\overline{\rho}}=&-a\overline{\rho} + \overline{R\rho} + b\overline{\rho^2}-c\overline{\rho^3}+h+\frac{\sigma}{L}\eta\,,\\
    \dot{\overline{R}}=&\delta - \frac{1}{\tau_D}(\overline{R\rho}+\frac{\sigma}{L}\zeta)\,.
\end{aligned}
\end{equation}

Note that the diffusion term vanishes when summed over the entire lattice with PBC, resulting in zero contribution. Additionally, the strength of the noise is reduced by a factor of $L$ due to the central limit theorem.

Using the relationships $\overline{R\rho}=\overline{R}\overline{\rho}+\mathrm{cov}(\rho, R)$, $\overline{\rho^2}=\overline{\rho}^2+\sigma_\rho^2$, $\overline{\rho^3}=\overline{\rho}^3+3\overline{\rho}\sigma_\rho^2+\gamma_\rho\sigma_\rho^3$, where $\sigma_\rho$ is the standard deviation and $\gamma_\rho$ is the skewness of $\rho$, the dynamics can be expressed as

\begin{equation}
\begin{aligned}
    \dot{\overline{\rho}}=&(-a+\overline{R}-3c\sigma_\rho^2)\overline{\rho} +  b\overline{\rho}^2-c\overline{\rho}^3+h+b\sigma_\rho^2-c\gamma_\rho\sigma_\rho^3\\
    &+\mathrm{cov}(\rho,R)+\frac{\sigma}{L}\eta\,,\\
    \dot{\overline{R}}=&\delta - \frac{1}{\tau_D}(\overline{R}\overline{\rho}+\mathrm{cov}(\rho,R)+\frac{\sigma}{L}\zeta)\,.
\end{aligned}
\end{equation}

Here, we introduce some approximations to simplify our understanding of the system. First, for large systems where $L\to\infty$, the terms $\frac{\sigma}{L}\eta$ and $\frac{\sigma}{L}\zeta$ approach 0. Additionally, numerical evidence from the down phase indicates that both $\rho_{\vec{x}}$ and $R_{\vec{x}}$ experience minimal fluctuations around their equilibrium values primarily due to the noise terms $\sigma\eta_{\vec{x}}$ and $\sigma\zeta_{\vec{x}}$. Consequently, we approximate $\mathrm{cov}(\rho,R)\approx 0$, and $\sigma_\rho^2\gg c\gamma_\rho\sigma_\rho^3\approx 0 $. Numerical simulations at $L=16$, $\tau_D=10$ provide values $\sigma_\rho=0.0353$, $\mathrm{cov}(\rho, R)=9.84\times10^{-5}$ and $\gamma_\rho=0.410$, supporting the validity of these approximations.  With these simplifications, the dynamics of the mean fields are described by

\begin{equation}
\begin{aligned}
    \dot{\overline{\rho}}=&(-a+\overline{R}-3c\sigma_\rho^2)\overline{\rho} +  b\overline{\rho}^2-c\overline{\rho}^3+h+b\sigma_\rho^2\,,\\
    \dot{\overline{R}}=&\delta - \frac{\overline{R}\overline{\rho}}{\tau_D}\,.
\end{aligned}\label{eq:mean}
\end{equation}

Next, we analytically estimate the value of $\sigma_\rho$. Assuming that, in the down phase, the fluctuation of $\rho_{\vec{x}}$ around its equilibrium follows a Gaussian distribution, $\Delta\rho_{\vec{x}}\sim N(0, \sigma_\rho^2)$. With a small fluctuation $\Delta\rho_{\vec{x}}$, we proceed by linearizing the dynamics of a single neural activity region, Eq.~\ref{dynamics_eqns1}, around its equilibrium point. At equilibrium, $(-a+R_{\vec{x}})\rho_{\vec{x}}+b\rho_{\vec{x}}^2-c\rho_{\vec{x}}^3+h=0$, and

\begin{figure*}[htbp]
    \centering
    \includegraphics[width=0.98\linewidth]{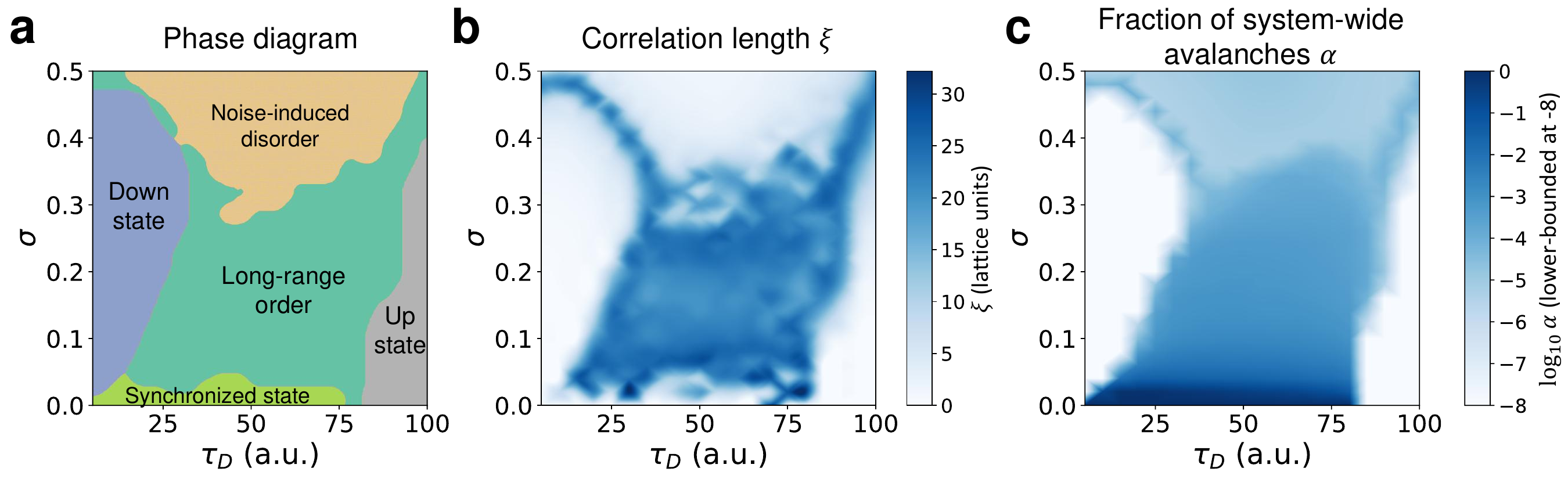}
    \caption{(a) Phase diagram of the collective dynamics, with $\tau_D$ and the noise strength $\sigma$ as variables, on a $64\times 64$ lattice. At small noise levels, all neural activity regions are synchronized, resulting in system-wide avalanches. An appropriate noise level breaks synchronization, and memory effects lead to LRO in the system, characterized by a power-law distribution of avalanches. As the noise level further increases, the system is dominated by noise, leading to a phase of noise-induced disorder. A down state is observed when $\tau_D$ is too small, and an up state is observed when $\tau_D$ is too large, both with short-range correlations. (b) The correlation length, $\xi$, defined as the average distance between two units within the same avalanche (excluding system-wide avalanches), as a function of $\sigma$ and $\tau_D$. An expansive phase of LRO with a large correlation length is observed. 
    (c): The fraction of system-wide avalanches, $\alpha$. In the synchronized state, all avalanches are system-wide, so $\alpha \to 1$. In both the down and up states, avalanche activity is absent, and $\alpha \to 0$. In the LRO phase, avalanches follow power-law distributions, leading to a small but finite value of $\alpha$. As the noise strength $\sigma$ increases, avalanche sizes diminish further, resulting in an even smaller—yet still nonzero—value of $\alpha$.}
    \label{fig:phase_diagram_noise}
\end{figure*}

\begin{equation}
\begin{aligned}
    \dot{\rho}_{\vec{x}}=&(-a+R_{\vec{x}})\Delta\rho_{\vec{x}}+2b\rho_{\vec{x}}\Delta\rho_{\vec{x}}-3c\rho_{\vec{x}}^2\Delta\rho_{\vec{x}}\\
    &+D\nabla^2(\Delta \rho_{\vec{x}})+\sigma\eta_{\vec{x}}\,.
\end{aligned}
\end{equation}

Taking a single discrete time step, $\Delta t$, we derive the following:

\begin{equation}
\begin{aligned}
     \Delta\rho_{\vec{x}}(t+\Delta t)&=\Delta\rho_{\vec{x}}(t) + \Big((R_{\vec{x}}-a+2b\rho_{\vec{x}}-3c\rho_{\vec{x}}^2)\Delta\rho_{\vec{x}}(t)\\
     +D&\sum_{\vec{y} \in n.n.\vec{x}}(\Delta\rho_{\vec{y}}(t)-\Delta\rho_{\vec{x}}(t))\Big)\Delta t + \sigma\eta_{\vec{x}}\sqrt{\Delta t}\,,\\
     &= \Delta\rho_{\vec{x}}(t)(1-(4D-\alpha_{\vec{x}})\Delta t) \\
     &\qquad +D\Delta t\sum_{\vec{y} \in n.n.\vec{x}}\Delta\rho_{\vec{y}}(t) + \sigma\eta_{\vec{x}}\sqrt{\Delta t}\,,
\end{aligned}
\end{equation}

\noindent where $\alpha_{\vec{x}}=R_{\vec{x}}-a+2b\rho_{\vec{x}}-3c\rho_{\vec{x}}^2$.

Assuming that $\Delta\rho_{\vec{x}}$ and $\Delta\rho_{\vec{y}}$ are uncorrelated and utilizing $\Delta\rho_{\vec{x}}\sim N(0, \sigma_\rho^2)$, we get:

\begin{equation}
\begin{aligned}
    N(0, \sigma_\rho^2)=&N(0, ((1-(4D-\alpha_{\vec{x}})\Delta t)\sigma_\rho)^2)\\
    &\quad + N(0, 4(D\Delta t\sigma_\rho)^2)+N(0, \sigma^2\Delta t),
\end{aligned}
\end{equation}

or

\begin{equation}
\begin{aligned}
    &\sigma_\rho^2=(1-(4D-\alpha_{\vec{x}})\Delta t)^2\sigma_\rho^2+4D^2\Delta t^2\sigma_\rho^2+\sigma^2\Delta t\,,\\
    \implies &\sigma_\rho^2=\frac{\sigma^2}{8D-2\alpha_{\vec{x}}-((4D-\alpha_{\vec{x}})^2+4D^2)\Delta t}\,.\label{eq:sigma_rho}
\end{aligned}
\end{equation}

Note that $\alpha_{\vec{x}}$ determines the stability of the equilibrium point in the single-region dynamics: $\alpha_{\vec{x}}>0$ indicates an unstable equilibrium, while $\alpha_{\vec{x}}<0$ denotes stability. Near the transition between the LRO phase and the down phase, $\alpha_{\vec{x}}$ is approximately 0, a finding supported by numerical results. In addition, with $\Delta t \ll D$,  the variance $\sigma_\rho^2$ simplifies to
\begin{equation}
    \sigma_\rho^2=\frac{\sigma^2}{8D}.
\end{equation}

After inserting constants, we get $\sigma_\rho=0.0354$, which closely aligns with our numerical results.

Now, the mean-field dynamics Eq.~\ref{eq:mean} incorporate all necessary variables, and an illustration of the nullclines is presented in Fig.~\ref{fig:nullcline}(d). By solving for the two turning points of the nullcline where $\dot{\overline{\rho}}=0$, we identify $\rho_1=0.0362$ and $\rho_2=0.748$, corresponding to the threshold values $\tau_D^\mathrm{low}=8.15$ and $\tau_D^\mathrm{high}=82.1$, respectively. The latter value is in agreement with the previous bound from Eq.~\ref{LROUpperBound} and our numerics.

Thus, we observe a down phase with no neural activities below $\tau_D=8.15$, while the lower bound noted in numerical experiments is $\tau_D \approx 25$. This discrepancy could be attributed to several factors: for instance, minor $\rho$ activities may fail to reach the predefined threshold $\rho_\mathrm{thrs}=0.5$ and go unrecorded in the numerical experiments; additionally, some assumptions, such as $\Delta\rho_{\vec{x}}\sim N(0, \sigma_\rho^2)$, may not hold at slightly larger $\tau_D$ values.

To further investigate the phase structure of the single-unit dynamics, we introduce an additional control parameter, $a$, the coefficient of the linear term in Eq.~\ref{dynamics_eqns1}. As shown in Fig.~\ref{fig:nullcline}(c), we identify a distinct phase boundary between spiking dynamics and the up state; however, the down state does not emerge. In contrast, when considering the mean-field equations, Eqs.~\ref{eq:mean}, a down state appears at small $\tau_D$ (Fig.~\ref{fig:nullcline}(f)), in agreement with our numerical experiments. Notably, the LRO phase observed in the main text spans nearly the entire spiking phase of the mean-field dynamics (Fig.~\ref{fig:nullcline}(f)), reinforcing that the LRO phase is well-separated from phase boundaries and cannot be attributed to critical effects.

\subsection{Phase Structure with Varying Noise Levels}
\label{sec:SI_phase_noise}

In the main text, we fixed the noise strength, $\sigma=0.1$. In this section, we demonstrate how varying the noise strength $\sigma$ changes the phase structure. 

When $\sigma\to 0$, the diffusion term in the activity dynamics, Eq.~\ref{dynamics_eqns1}, quickly leads to synchronized spiking behaviors. This results in the prevalence of system-wide avalanches, leading to a synchronized state. On the other hand, when $\sigma$ is too large, noise dominates the dynamics and leads to frequent threshold crossings, resulting in a high degree of uncorrelated neural activity that we call the noise-induced disordered state.

\begin{figure*}[htbp]
    \centering
    \includegraphics[width=\linewidth]{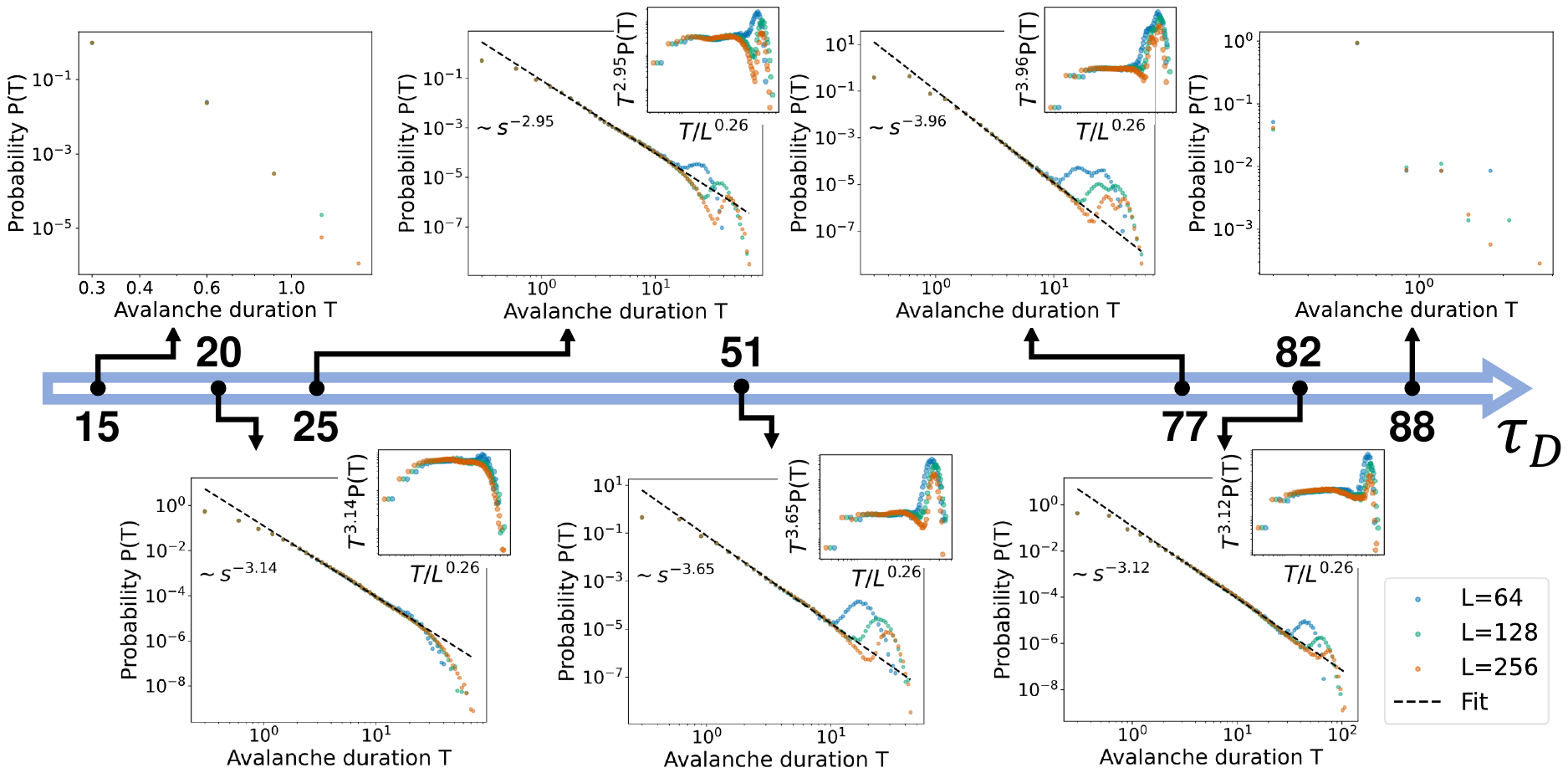}
\caption{Illustration of avalanche duration distributions for different values of $\tau_D$. Similar to Fig.~\ref{fig:FIG_3} in the main text, we see a down phase and a rigid phase with very short avalanche durations, and an LRO phase with avalanche durations following a power-law distribution. The insets show the curves rescaled according to the finite-size scaling ansatz, Eq.~\ref{decaying_power_law_distribution}, with $\beta_T=0.26$. For $L=\{64, 128, 256\}$, the statistics from each ensemble are taken from $\{100, 400, 2000\}$ instances, with each instance simulated for $\{50, 20, 5\} \times 10^{4}$ timesteps, respectively. Parameter values: $a = 1$, $b = 1.5$, $c = 1$, $h = 10^{-7}$, $D = 1$, $\sigma = 0.1$, $\delta = 0.004$, $\epsilon = 0.5$, $\Delta t=0.01$, and $\Delta_{tw}= 0.3$. 
    }
    \label{fig:big_size_fig_SI}
\end{figure*}

In Fig.~\ref{fig:phase_diagram_noise}(a), we computed the avalanche size distributions with varying $\tau_D$ and $\sigma$, on a $64\times 64$ lattice. A point belongs to the LRO phase when its avalanche sizes follow a power-law distribution of at least 3 decades, with a power-law exponent $\alpha_s$ between 1.5 and 2.5. We identify the synchronized phase by its system-wide avalanches, and the noise-induced disordered phase by its frequently active (yet uncorrelated) neural activities. The up and down phases are identified by their lack of avalanche activities altogether. With this analysis, we obtained the phase diagram in Fig.~\ref{fig:phase_diagram_noise}(a) (phase boundaries are smoothed from a grid using mode smoothing). In Fig.~\ref{fig:phase_diagram_noise}(b), we computed the correlation length $\xi$, defined as the average distance between two sites within the same avalanche (excluding system-wide avalanches), of varying parameter combinations; and in Fig.~\ref{fig:phase_diagram_noise}(c), we compute the fraction of system-wide avalanches, $\alpha$. The results further supports the validity of the phase diagram.  

To further investigate the role of noise, we remove the noise term, $\sigma\zeta_{\vec{x}}$, from the resource equation, Eq.~\ref{dynamics_eqns2}, and perform the same analysis as in the main text. The results, shown in Fig.~\ref{fig:resource_noise}, indicate that the outcomes remain qualitatively unchanged with or without the noise term in Eq.~\ref{dynamics_eqns2}. This confirms that noise in the resource dynamics is not the primary driver of the observed LRO phase, thereby ruling out the possibility of a temporal Griffiths phase \cite{moretti2013griffiths, vazquez2011temporal}.

\begin{figure*}[htbp]
    \centering    \includegraphics[width=0.7\linewidth]{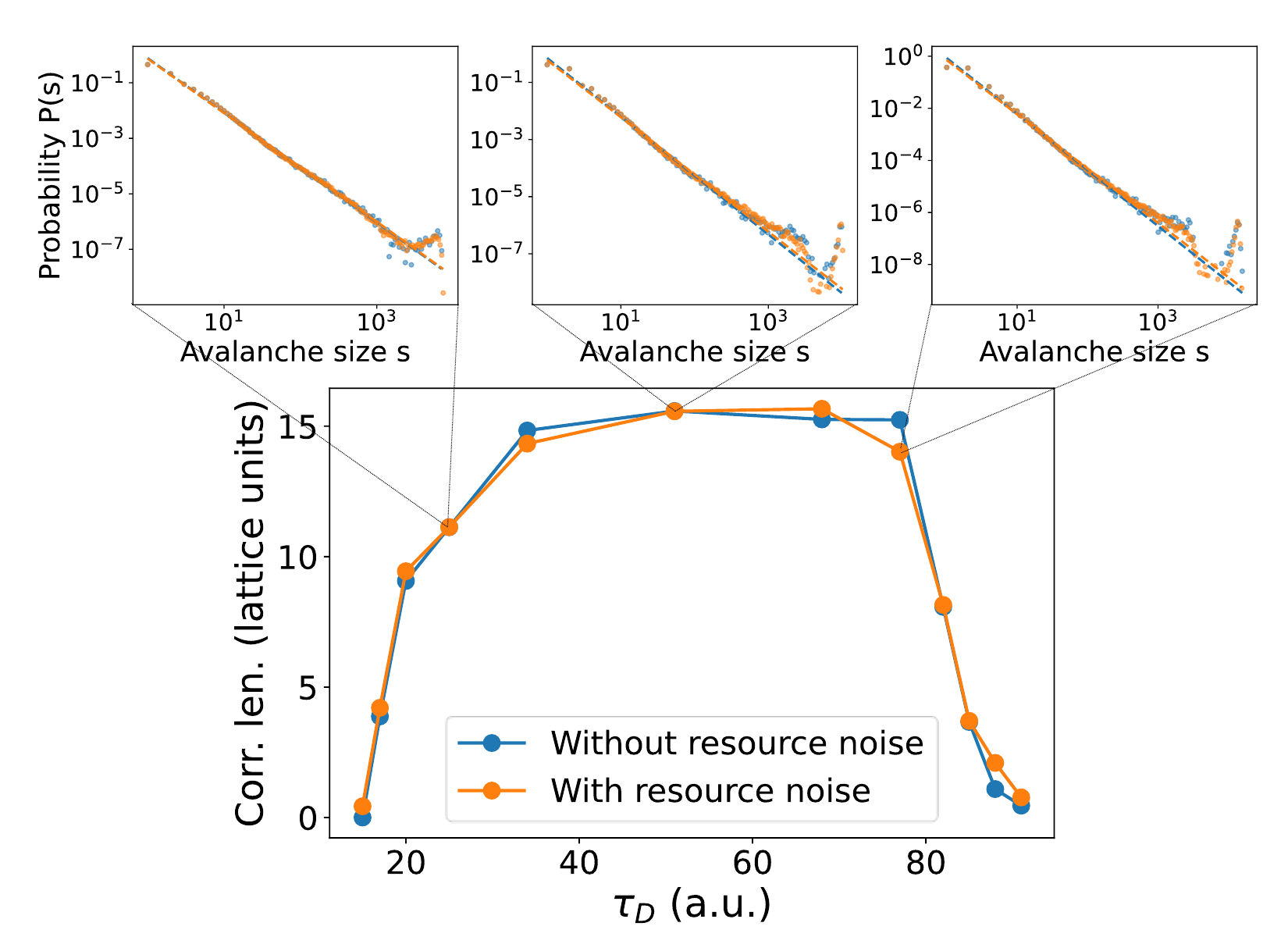}
\caption{Comparison of avalanche size distributions and correlation lengths with and without the noise term in the resource equation, Eq.~\ref{dynamics_eqns2}. In this numerical experiment, we set $L = 64$ and $\sigma = 0.1$. The results show that the two scenarios produce nearly identical outcomes, with overlapping curves. This indicates that the noise term in the resource equation does not contribute to the emergence of the LRO phase, further ruling out the possibility that the latter originates from a Griffiths phase.}
    \label{fig:resource_noise}
\end{figure*}

\subsection{Temporal Avalanche Distributions}
\label{sec:SI_size_distribution}

Fig.~\ref{fig:FIG_3} in the main text shows avalanche size ($s$) distributions.  In a similar manner, we define the avalanche duration, $T$, as the time interval from the start to the end of each avalanche. Fig.~\ref{fig:big_size_fig_SI} shows the distribution of avalanche durations for $\tau_D\in\{15, 20, 25, 51, 77, 82, 88\}$. Consistent with Fig.~\ref{fig:FIG_3} in the main text, we find power-law distributions of avalanche durations within the LRO phase, whereas durations are notably shorter in both the down and rigid phases. Using the finite-size scaling ansatz, Eq.~\ref{decaying_power_law_distribution}, the insets of Fig.~\ref{fig:big_size_fig_SI} show that the rescaled distribution curves approximately overlap with each other.

\begin{figure}[htbp]
    \centering
    \includegraphics[width=\linewidth]{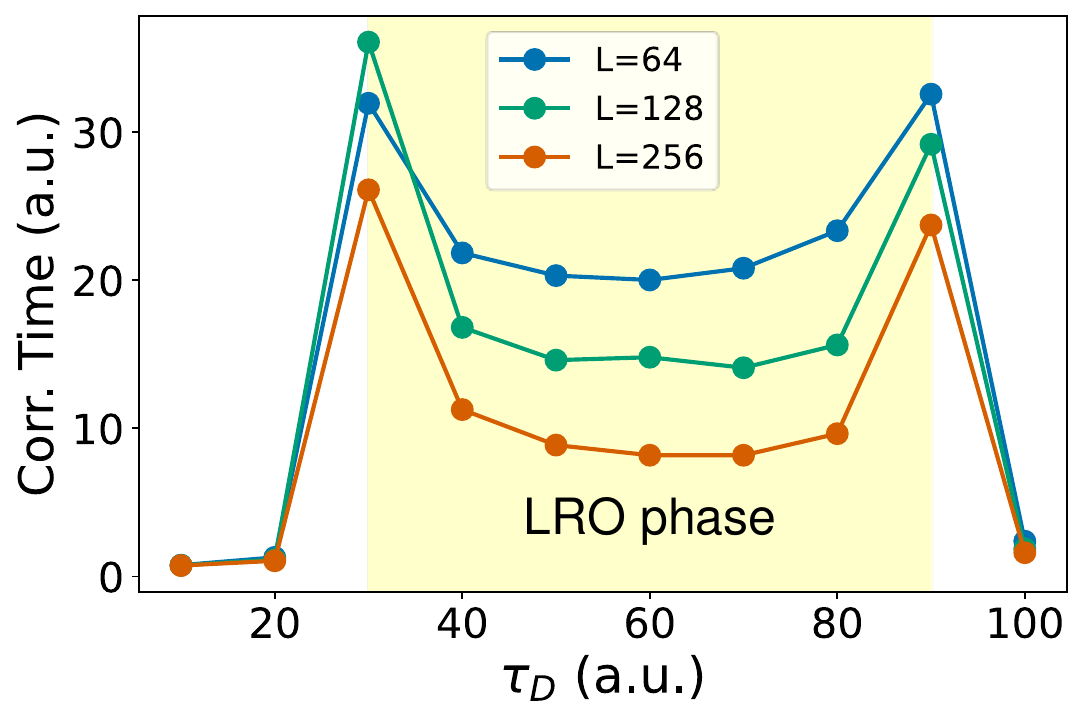}
    \caption{The correlation time, defined as the average distance in time between two events within the same avalanche, as a function of $\tau_D$. We observe an LRO phase with long correlation times, contrasting the down phase and the rigid phase with correlation times approaching 0. For better visualization, the noise strength is $\sigma=0.22$ in this figure instead of $0.1$ in the main text, so that the synchronization effect is weaker and we see longer avalanches. As a consequence, the LRO phase slightly shifts to $30 \le \tau_D \le 90$.}
    \label{fig:correlation_time}
\end{figure}

Similar to the correlation length discussed in the main text, we define the correlation time as the average distance in time between two events within the same avalanche. The correlation time as a function of $\tau_D$ is depicted in Fig.~\ref{fig:correlation_time}. We excluded the case where $L=512$ due to the longest avalanches potentially exceeding the simulation duration. We observe extended correlation times indicative of an LRO phase, contrasting sharply with the negligible correlation times of the down and rigid phases. Notably, the correlation times near the boundaries of the LRO phase are the most extended and may potentially diverge. The emergence of the two peaks in the correlation time plot at the boundary with the ``up'' and ``down'' phases is related to larger fluctuations and ``smearing'' of the activity waves near the phase boundaries. As a consequence, the activity waves propagate slower at the boundaries, leading to longer avalanches. In comparison, in the middle of the LRO phase, the activity waves are more confined and propagate faster. However, due to limited computational resources, we could not precisely determine the phase transition points, leaving this for future investigation.


\subsection{Scaling Relation Analysis}
\label{sec:scaling_relations}

\begin{figure*}[htbp]
    \centering
    \includegraphics[width=\linewidth]{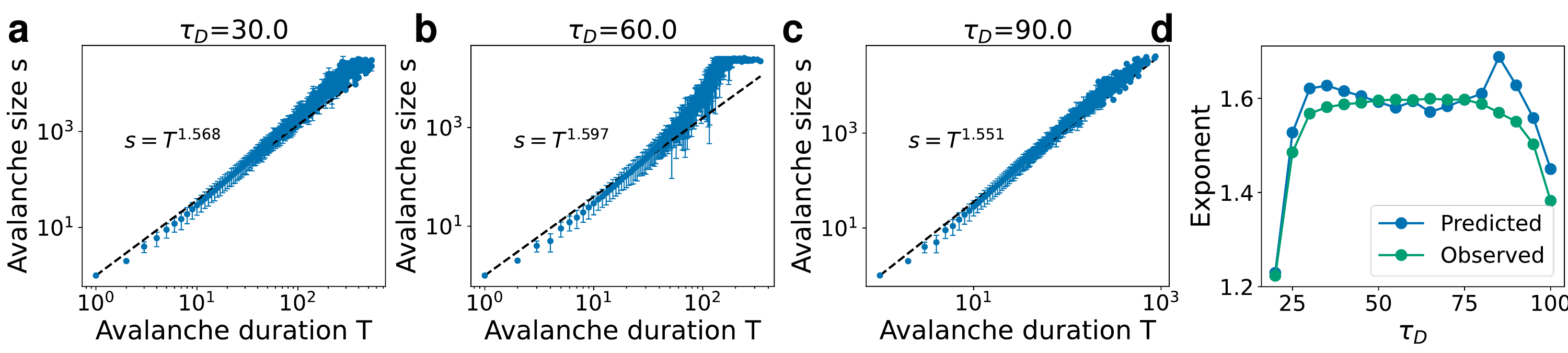}
    \caption{(a-c) The mean avalanche size $\langle s \rangle$ as a function of the avalanche duration $T$, for $\tau_D=\{30, 60, 90\}$. In these numerical tests, the lattice size was set to $L=64$, and the noise strength $\sigma=0.22$. Error bars represent one standard deviation. Their relation can be fitted with a power-law, $\langle s \rangle = T^\gamma$. (d) Comparison of the predicted exponent, $\gamma=\frac{\alpha_T-1}{\alpha_s-1}$, and the numerically fitted exponents as a function of $\tau_D$. The numerical results agree well with the theoretical predictions.} 
    \label{fig:size_vs_duration}
\end{figure*}

To derive the scaling relation Eq.~\ref{eq:scaling_eqn}, we must start with the probability distribution associated with having an avalanche of size $s$ {\it and} duration $T$ (assuming we are in the LRO phase)~\cite{sethna2005randomfieldisingmodelshysteresis}:

\begin{equation}
    P(s, T) \propto s^{-(\tau + \sigma \nu z)} \overline{P} \bigg( \frac{s}{s_s}r^{1/\sigma}, \frac{T}{T_s}r^{\nu z} \bigg)\,.
    \label{eq:general_avalanche_distrb}
\end{equation}

Here, we have integrated out additional variables from~\cite{sethna2005randomfieldisingmodelshysteresis}, as our avalanches are characterized by size $s$ and duration $T$ alone (plus some additional dependence on exponents through $r$, the disorder; again, see~\cite{sethna2005randomfieldisingmodelshysteresis} for details). $\overline{P}$ is a scaling function, $s_s$ and $T_s$ are some characteristic size and time scales, and $\tau$, $\sigma$, $\nu$, and $z$ are all exponents which we will soon relate to the exponents in Eq.~\ref{eq:scaling_eqn}. From further integrating out $T$, we can determine a scaling relation for $P(s)$:

\begin{equation}
\begin{aligned}
    P(s) &= \int dT \, P(s, T) \\
    &\propto \int dT \, s^{-(\tau + \sigma \nu z)} \overline{P} \bigg( \frac{s}{s_s}r^{1/\sigma}, \frac{T}{T_s}r^{\nu z} \bigg) \\
    &\propto \int dx \, r^{- \nu z} s^{-(\tau + \sigma \nu z)} \overline{P} \bigg( \frac{s}{s_s}r^{1/\sigma}, x \bigg) \\
    &\propto s^{-\tau} \int dx \, \bigg( \frac{s}{s_s}r^{1/\sigma} \bigg)^{- \sigma \nu z} \overline{P} \bigg( \frac{s}{s_s}r^{1/\sigma}, x \bigg) \\
    &\propto s^{-\tau} \overline{P}^*_1 \bigg( \frac{s}{s_s}r^{1/\sigma} \bigg)\,.
    \label{eq:P(s)_derivation}
\end{aligned}
\end{equation}

Above, we performed a change of variables and then absorbed the remaining factors of $\frac{s}{s_s}r^{1/\sigma}$ into a new scaling function $\overline{P}^*_1$. We are left with a function proportional to $s^{-\tau}$, so we recognize $\tau = \alpha_s$ from Sec.~\ref{sec:numerics}. We find $P(T)$ similarly:

\begin{equation}
\begin{aligned}
    P(T) &=  \int ds \, P(s, T) \\
    &\propto \int ds \, s^{-(\tau + \sigma \nu z)} \overline{P} \bigg( \frac{s}{s_s}r^{1/\sigma}, \frac{T}{T_s}r^{\nu z} \bigg) \\
    &\propto \int dy \, r^{-1/\sigma} (y r^{-1/\sigma})^{-(\tau + \sigma \nu z)} \overline{P} \bigg( y, \frac{T}{T_s}r^{\nu z} \bigg) \\
    &\propto T^{-(1 + \frac{\tau - 1}{\sigma \nu z})} \int dy \, y^{-(\tau + \sigma \nu z)} \\
    &\qquad \qquad \times \bigg( \frac{T}{T_s}r^{\nu z} \bigg)^{(1 + \frac{\tau - 1}{\sigma \nu z})} \overline{P} \bigg( y, \frac{T}{T_s}r^{\nu z} \bigg) \\
    &\propto T^{-(1 + \frac{\tau - 1}{\sigma \nu z})} \overline{P}^*_2 \bigg( \frac{T}{T_s}r^{\nu z} \bigg)\,.
    \label{eq:P(T)_derivation}
\end{aligned}
\end{equation}

This gives us $\alpha_T = 1 + \frac{\tau - 1}{\sigma \nu z}$, in line with Sec.~\ref{sec:numerics}. Lastly, we find $\langle s \rangle (T)$ from the conditional probability distribution $P(s | T) \equiv P(s, T)/P(T)$:

\begin{equation}
\begin{aligned}
    \langle s \rangle (T) &= \int ds \, s P(s | T) \\ 
    &= \int ds \, s \frac{P(s, T)}{P(T)} \\
    &\propto \int ds \, s^{1 - (\tau + \sigma \nu z)} T^{(1 + \frac{\tau - 1}{\sigma \nu z})} \overline{Q}\bigg( \frac{s}{s_s}r^{1/\sigma}, \frac{T}{T_s}r^{\nu z} \bigg) \\
    &\propto \int dy \, r^{-1/\sigma} (y r^{-1/\sigma})^{1 - (\tau + \sigma \nu z)} \\
    &\qquad \qquad \times T^{(1 + \frac{\tau - 1}{\sigma \nu z})} \overline{Q}\bigg( y, \frac{T}{T_s}r^{\nu z} \bigg) \\
    &\propto T^{1/\sigma \nu z} \int dy \, y^{1 - (\tau + \sigma \nu z)} \\
    &\qquad \qquad \times \bigg( \frac{T}{T_s}r^{\nu z} \bigg)^{(1 + \frac{\tau - 2}{\sigma \nu z})} \overline{Q}\bigg( y, \frac{T}{T_s}r^{\nu z} \bigg) \\
    &\propto T^{1/\sigma \nu z} \overline{Q}^*\bigg( y, \frac{T}{T_s}r^{\nu z} \bigg) \,,
    \label{eq:s(T)_derivation}
\end{aligned}
\end{equation}

where $\overline{Q}$ and $\overline{Q}^*$ are some new scaling functions defined in terms of previous ones. It is easy to then confirm that Eq.~\ref{eq:scaling_eqn} holds, where $\alpha_s = \tau$, $\alpha_T = 1 + \frac{\tau - 1}{\sigma \nu z}$, and $\gamma = 1/\sigma \nu z$:

\begin{equation}
\begin{aligned}
    \gamma = \frac{\alpha_T - 1}{\alpha_s - 1} \iff \frac{1}{\sigma \nu z} = \frac{1 + \frac{\tau - 1}{\sigma \nu z} - 1}{\tau - 1}\,.
\end{aligned}
\end{equation}

In Fig.~\ref{fig:size_vs_duration}, we plot $\langle s \rangle (T)$ for various $\tau_D$ (within the LRO phase) and compare theoretical predictions with numerical fits. We find that they agree reasonably well for all $\tau_D$. Since our distributions abide by finite-size scaling, we can attribute the bumps in Fig.~\ref{fig:size_vs_duration} (a-c) to finite-size effects.

\begin{figure*}[htbp]
    \centering
    \includegraphics[width=\linewidth]{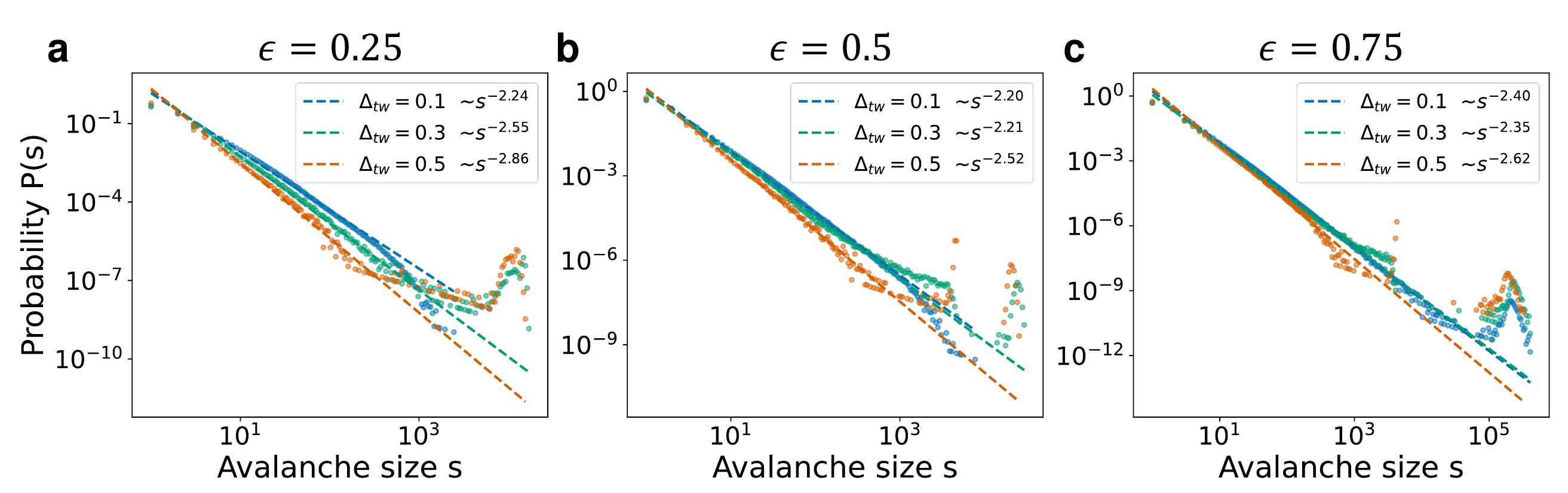}
    \caption{Avalanche size distributions for different time windows, $\Delta_{tw} = {0.1, 0.3, 0.5}$, and thresholds, $\epsilon = {0.25, 0.5, 0.75}$, at $\tau_D = 82$, $\sigma = 0.1$, and $L = 64$. The distributions show only weak dependence on these parameters, confirming the robustness of our avalanche definition.} \label{fig:window_threshold}
\end{figure*}

\begin{figure*}[htbp]
    \centering
    \includegraphics[width=\linewidth]{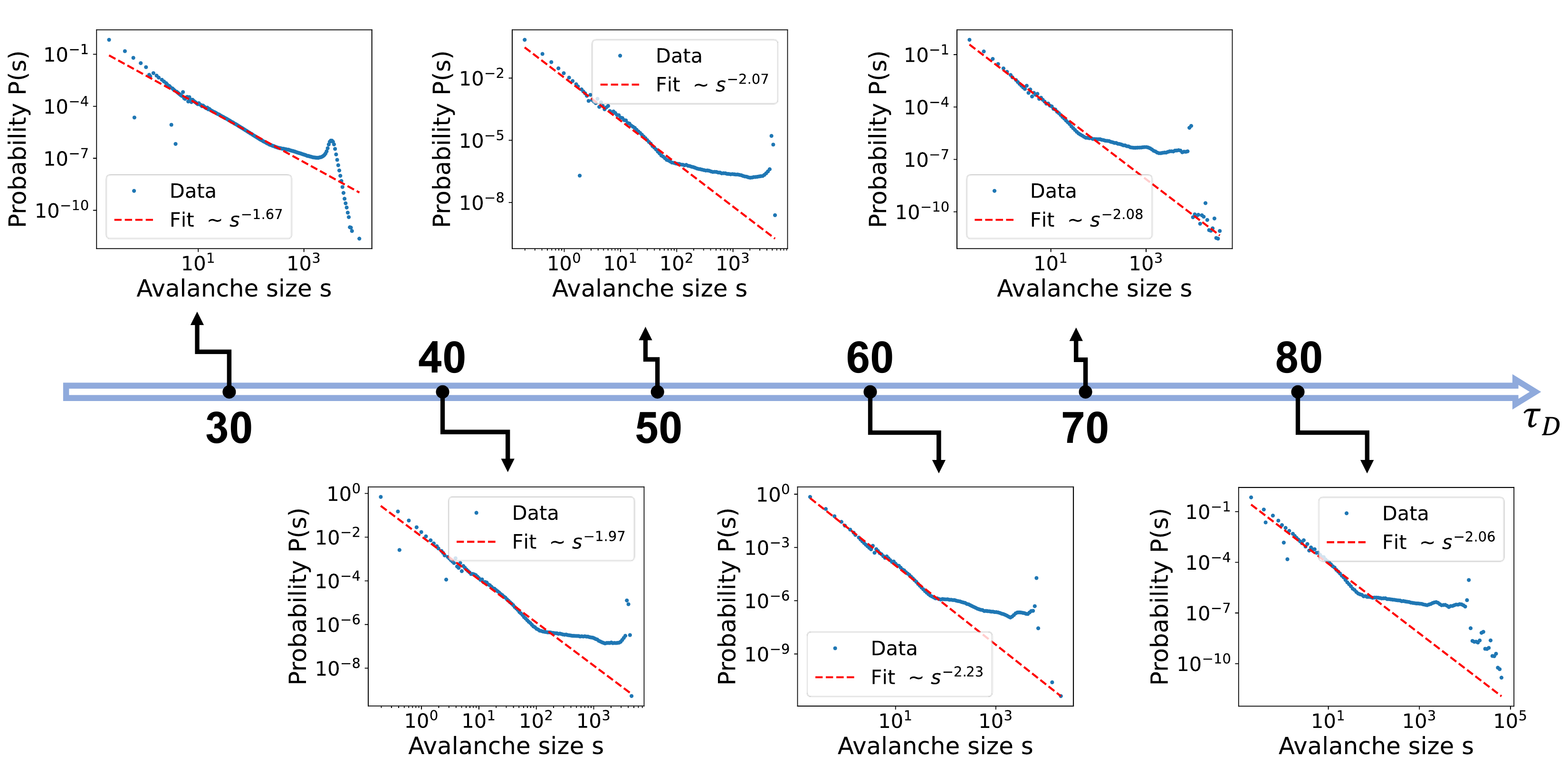}
    \caption{Avalanche size distribution using an alternative definition, where the average neural activity exceeding a threshold ($\varepsilon = 0.2$) is integrated over time, without considering the spatial structure of the neural activity. Avalanches are computed on a $64\times 64$ lattice, with noise strength $\sigma=0.22$. Using this definition, we observe a similar LRO phase, confirming that our main results are robust with respect to the specific choice of avalanche definition.}
    \label{fig:avalanche_definition}
\end{figure*}

\subsection{Avalanche Definition and Dependence on Numerical Parameters}

In this work, an avalanche is defined as a sequence of threshold-crossing events in neural activity that occur in close spatial and temporal proximity. In the main text, we set the threshold $\epsilon = 0.5$, define spatial proximity as one lattice unit, and use a temporal proximity window of $\Delta_{tw} = 0.3$. While these parameter choices are somewhat arbitrary, we demonstrate here that the extracted avalanche statistics exhibit only weak dependence on them.

To assess this, we computed avalanche size distributions for varying thresholds $\epsilon$ and time windows $\Delta_{tw}$, with the results shown in Fig.~\ref{fig:window_threshold}. While modifying these parameters affects the tails of the distributions, the power-law behavior remains intact, with only slight variations in the power-law exponents. This confirms that our avalanche definition is robust and largely insensitive to numerical parameter choices.

To compare with existing literature, we also tested an alternative definition of avalanches commonly used in experimental settings \cite{munoz_paper, Beggs_and_Plenz_avalanche}. In this approach, an avalanche begins when the average neural activity exceeds a threshold, $\varepsilon$, and ends when the average neural activity falls below $\varepsilon$. The avalanche size is defined as the average neural activity integrated over the duration of the avalanche. This definition does not rely on knowledge of the spatial structure of the network and is more readily applicable in experimental contexts.

In Fig.~\ref{fig:avalanche_definition}, we show the avalanche size distributions computed using this alternative definition with $\varepsilon = 0.2$, for various values of $\tau_D$, with system size $L=64$, noise strength $\sigma=0.22$. The range over which LRO appears remains similar, and the distributions consist of a leading power-law segment followed by a tail of system-wide avalanches. Additionally, we observe a sharper transition between phases: no avalanches occur for $\tau_D < 28$, while for $\tau_D > 88$ a single persistent avalanche dominates (not shown in Fig.~\ref{fig:avalanche_definition}, as it does not yield a distribution) Despite these differences, this alternative definition reinforces our main conclusion—the existence and extent of the LRO phase do not depend on the specific definition of avalanches.

\subsection{Synchronization Transition}

\begin{figure}[htbp]
    \centering
    \includegraphics[width=\linewidth]{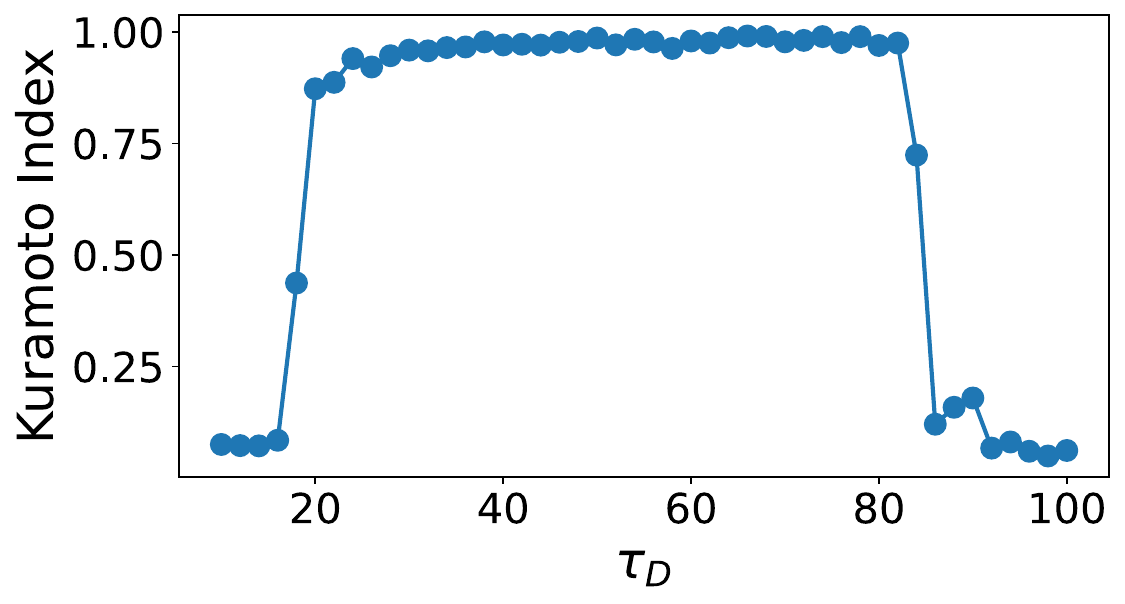}
    \caption{Kuramoto index $K$ as a function of $\tau_D$, computed on a $64 \times 64$ lattice with noise strength $\sigma = 0.1$. A clear partial synchronization transition is observed: for $20 \le \tau_D \le 82$, the system exhibits high phase coherence ($K\sim 0.9$), corresponding to the LRO phase identified in the main text. Outside this range—when $\tau_D < 20$ or $\tau_D > 82$—the system becomes desynchronized, with $K$ approaching zero.}
    \label{fig:kuramoto_index}
\end{figure}

To further characterize the nature of the LRO phase, we quantify neuronal synchronization using the Kuramoto index $K$ \cite{strogatz2003synchronization, munoz_paper}, defined as
\begin{equation}
    K=\frac{1}{N}\Big\langle \Big| \sum_{\vec{x}} e^{i\phi_{\vec{x}}(t)} \Big| \Big\rangle
\end{equation}
where $N$ is the number of neurons and $\langle \cdot \rangle$ denotes a combined time and ensemble average. Here, $\phi_{\vec{x}}(t)$ is the instantaneous phase of the neuron at site $\vec{x}$. A value of $K = 1$ indicates full synchronization, while $K = 0$ corresponds to completely desynchronized dynamics.

In our system, the phase $\phi_{\vec{x}}(t)$ is extracted via the Hilbert transform of the neural activity $\rho_{\vec{x}}(t)$:
\begin{equation}
\begin{aligned}
    \tilde{\rho}_{\vec{x}}^{\text{real}}(t)=&\rho_{\vec{x}}(t)-\langle\rho_{\vec{x}}\rangle,\\
    \tilde{\rho}_{\vec{x}}^{\text{imag}}(t)=&\mathcal{H}[\tilde{\rho}_{\vec{x}}^{\text{real}}(t)],\\
    \phi_{\vec{x}}(t)=&\arctan\bigg(\frac{\tilde{\rho}_{\vec{x}}^{\text{imag}}(t)}{\tilde{\rho}_{\vec{x}}^{\text{real}}(t)}\bigg).
\end{aligned}
\end{equation}
where $\mathcal{H}[\cdot]$ denotes the Hilbert transform.

Fig.~\ref{fig:kuramoto_index} shows the computed Kuramoto index as a function of $\tau_D$. Two sharp transitions are evident at $\tau_D = 20$ and $\tau_D = 82$. Within the range $20 \le \tau_D \le 82$, $K$ remains close to 1, yet strictly below it, indicating strong global phase coherence without perfect synchrony. This range coincides with the LRO phase identified in the main text. 

The fact that $K$ is close to 1 reflects the fact that neural activity across the lattice is approximately periodic and broadly aligned in phase: most sites fire in a repeating, lattice-wide rhythm. However, $K$ does not reach 1 because the firing times are not exactly identical across all sites. Instead, there is a small but spatially structured timing offset from site to site within each oscillation cycle. These timing offsets are large enough to break each cycle into multiple spatiotemporal activation clusters (the avalanches we analyze), rather than a single perfectly simultaneous burst, but still small compared to the overall oscillation period. As a result, the system exhibits both (i) near-synchronous oscillations at the global level and (ii) scale-free avalanche statistics at the level of individual co-activation events.

Outside this range, in the down phase ($\tau_D < 20$) and the up phase ($\tau_D > 82$), $K$ drops to near zero. In these regimes, neural activity is either absent or static, and oscillatory dynamics are suppressed, leading to negligible phase coherence.

\renewcommand{\figurename}{Supplementary Movie}
\setcounter{figure}{0}

\begin{figure*}[htbp]
    \centering
    \includegraphics[width=\linewidth]{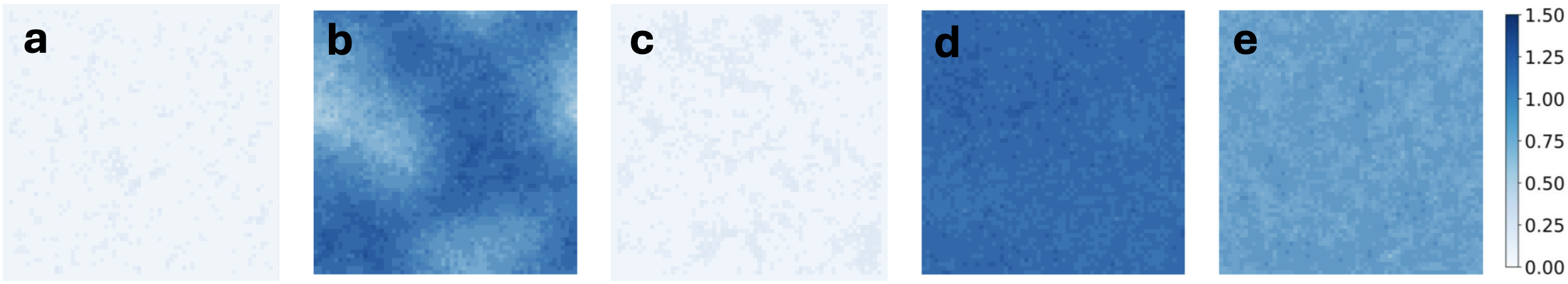}
    \caption{Dynamics of neural activities $\rho$ on a lattice of size $L^2 = 64^2$, with noise level $\sigma=0.1$. Darker blues correspond to higher activities. (a) Dynamics for $\tau_D = 15$, in the down phase. In this regime, there is no activity except in small bursts due to the presence of noise. (b-d) Dynamics for $\tau_D = 25$, $51$, and $77$, respectively, corresponding to the LRO phase. Here, large-scale, highly correlated waves of activity can be observed. (e) Dynamics for $\tau_D = 88$. When $\tau_D$ is sufficiently large ($\tau_D \gtrsim 82$), the activities remain at high values perpetually since the resource decay timescale is extremely slow.} 
\end{figure*}

\begin{figure*}[htbp]
    \centering
    \includegraphics[width=\linewidth]{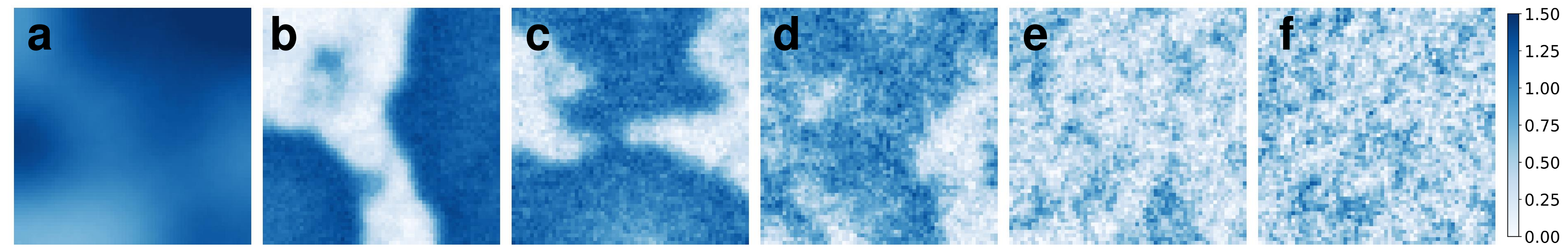}
    \caption{Dynamics of neural activities $\rho$ on a lattice of size $L^2 = 64^2$, with $\tau_D=51$. Darker blues correspond to higher activities. (a-f) Dynamics for noise levels $\sigma=\{0, 0.1, 0.2, 0.3, 0.4, 0.5\}$. (a) When $\sigma=0$, all neural activities in the system are synchronized, and avalanches always span the entire system. (b-d) When $\sigma\in\{0.1, 0.2, 0.3\}$, we see correlated waves of neural activity, and avalanches follow a power-law distribution. (e-f) When $\sigma\in\{0.4, 0.5\}$, most activities in the system are induced by noise, and the random flipping events lead to persistent, noise-induced avalanches that lasts indefinitely. }
\end{figure*}

\begin{figure*}[htbp]
    \centering
    \includegraphics[width=\linewidth]{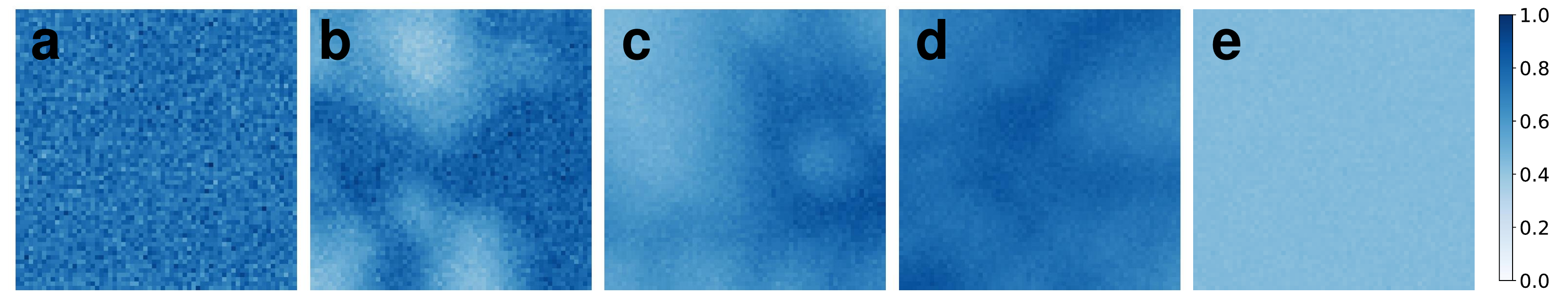}
    \caption{Dynamics of synaptic resources $R$ on a lattice of size $L^2 = 64^2$, with noise level $\sigma = 0.1$. Darker blues correspond to higher resource levels. (a) $\tau_D = 15$ (down phase): resource levels remain at an intermediate value, insufficient to sustain neural activity. (b–d) $\tau_D = 25$, $51$, and $77$ (LRO phase): waves of rising and falling resource levels are clearly observed. (e) $\tau_D = 88$ (up phase): with persistent neural activity, resource levels stabilize near a constant value.}
\end{figure*}

\end{document}